**Grain boundaries amplify local chemical ordering in complex concentrated alloys**

Ian Geiger [1], Yuan Tian [2], Ying Han [2], Yutong Bi [2], Xiaoqing Pan [2], Penghui Cao [2,3], Timothy J. Rupert [1,2,4,5,*]

[1] Materials and Manufacturing Technology, University of California, Irvine, CA 92697, USA

[2] Department of Materials Science and Engineering, University of California, Irvine, CA 92697, USA

[3] Department of Mechanical and Aerospace Engineering, University of California, Irvine, CA 92697, USA

[4] Hopkins Extreme Materials Institute, Johns Hopkins University, Baltimore, MD 21218, USA

[5] Department of Materials Science and Engineering, Johns Hopkins University, Baltimore, MD 21218, USA

* Corresponding author: tim.rupert@jhu.edu

## ABSTRACT

Local chemical ordering strongly influences the behavior of complex concentrated alloys, yet its characterization remains challenging due to the nanoscale dimensions and scattered spatial distribution of the ordered domains. Here, we study chemical ordering near grain boundaries, demonstrating that interfaces can act as microstructural anchor points that amplify chemical order and promote near-boundary compositional modulation. Atomistic simulations reveal the development of composition waves with ordering vectors normal to the boundary plane in two distinct material systems, CrCoNi and NbMoTaW. These waves manifest as periodic enrichment-depletion patterns that reflect the underlying chemical ordering tendencies of each system, with amplified contrast that can extend several nanometers into the grain interior before gradually decaying. By examining multiple grain boundary orientations and alloys, we show that both the interfacial segregation profile and the crystallographic terminating plane govern the extent and character of this amplification. This interplay between boundary-dictated directional ordering and the diffuse, untemplated chemical domain evolution within the grain advances our understanding of interface-mediated ordering phenomena and suggests new opportunities for experimentally detecting local chemical order in complex concentrated alloys.

## I. Introduction

Complex concentrated alloys (CCAs) are a novel class of alloys formed by mixing three or more elements in roughly equimolar proportions [1]. A salient feature of these alloys is the diverse chemical interactions that govern their local atomic configurations. In CCAs with strongly interacting elements, selective bonding or repulsion between species drives the formation of nanoscale chemical domains typically referred to as local chemical ordering (LCO) [2]. LCO can be further classified by scale: chemical short-range order (CSRO) involves correlations within ~1 nm [3–5], while chemical medium-range order extends over larger distances (~1-5 nm) [6]. Unlike long-range periodic order, LCO roughens the potential energy surface, influencing the energetics and diffusion pathways of defects such as vacancies [7–9], dislocations [4,10,11], and grain boundaries [8,12–14]. For example, dislocations in CCAs can adopt wavy morphologies due to the heterogeneous atomic environments that impede their motion, enhancing mechanical performance [15–17]. At grain boundaries, complex segregation states, sometimes driven by co-segregation [18,19], modify interfacial structures and properties. These states can affect structural transformation pathways [14,20,21], suppress grain growth [22–25], and promote second-phase nucleation [25]. The variability of boundary structures and compositions in CCAs makes these effects difficult to predict, highlighting the need to clarify the mechanisms driving chemical order at grain boundaries.

Atomistic simulations are well-suited for exploring these mechanisms because they provide spatial and chemical correlations on the relevant length scales for LCO. For example, density functional theory (DFT) calculations have shown that frustrated magnetic alignments in CrCoNi minimize like-spin Cr bonds [26], consistent with extended X-ray absorption fine structure (EXAFS) [27] analyses revealing preferential Cr-Ni and Cr-Co bonding. Alternatively, hybrid

Monte Carlo (MC)/molecular dynamics (MD) simulations capture the effect of compositional complexity on microstructural evolution mechanisms such as twinning [28], phase transformations [29], and grain boundary state transitions [14,20] in larger systems across varied temperatures and timescales. These methods reveal chemical order transitions in both face-centered cubic (FCC) [15,30] and body-centered cubic (BCC) CCAs [31–33], with lower temperatures generally enhancing LCO. Furthermore, segregation near defects often strengthens chemical pair correlations [13,32,34], linking structural disorder to concentrated chemical ordering.

At the same time, experimental identification of LCO in CCAs remains a challenge due to the exceedingly small size of clusters. These features often have domain widths on the order of a nanometer, making characterization difficult within the crystalline matrix. EXAFS measurements by Fantin et al. [35] on BCC NbMoTaW revealed high-temperature ordering signatures consistent with the B2-type (Mo-Ta) and B32-type (Nb-W) ordering predicted at lower temperatures. Advanced transmission electron microscopy (TEM) techniques have provided complementary insight into these short-range ordered structures. For example, scanning TEM high-angle annular dark field (STEM-HAADF) imaging leverages Z-contrast sensitivity to detect LCO when regions are enriched with elements of differing atomic numbers, as shown by Lei et al. [36] in a BCC TiZrHfNb alloy. Selected area electron diffraction (SAED) patterns showing extra superlattice reflections are another widely used indicator of LCO [37]. In addition, streaking and diffuse discs at non-Bragg spots in reciprocal space can signal spatial ordering under appropriate zone-axis conditions [4,37–41]. For example, nanobeam diffraction has revealed extra diffuse disks at ½$\{\bar{3}11\}$ positions along the [112] zone axis, proposed as evidence of CSRO in face-centered cubic (FCC) VCoNi [4,42], CrCoNi [5], and $Fe_{50}Mn_{30}Co_{10}Cr_{10}$ [40]. However, the origin of these types of reflections remain a subject of debate [43–45], as similar features are observed in pure metals [46] and can also result from structural imperfections [47]. Moreover, scattering between reciprocal lattice planes from higher-order Laue zone diffraction can produce diffuse features unrelated to local composition along specific zone axes [44,46]. This ongoing uncertainty highlights the limitations of current experimental methods and underscores the gap between computational and experimental evidence of LCO.

To address this gap, new strategies for predictably localizing LCO within microstructures should be explored. Grain boundaries are promising in this context, as their structure and composition have been shown to drive other types of chemical transitions in nearby regions. For example, coherency stresses associated with nanoscale structure formation can be relieved at grain boundaries and surfaces [48,49], making these defects natural pinning centers for chemical segregation and patterning. Extensive phase-field studies of spinodal decomposition in polycrystalline models show that attenuating concentration bands form parallel to interfaces due to segregation, driven by a balance between grain boundary-assisted phase separation and normal spinodal decomposition within the grain [48,50,51]. The high internal lattice stresses in some CCAs, caused by LCO-related symmetry disruption and complex sublattice occupation [52–54], suggest that similar mechanisms may be active in CCAs with strong alloy interactions and lattice mismatch between ordered structures. Moreover, preferential segregation at grain boundaries can locally reduce chemical complexity, potentially creating enriched regions that promote co-segregation in nearby crystalline sites via the strong, intrinsic interactions that drive LCO formation.

In this study, atomistic simulations are used to study LCO and compositional patterning near a Σ11 boundary in CrCoNi and a series of Σ3 boundaries in NbMoTaW. Both alloys show strong ordering tendencies [5,35,38,55,56], with Han et al. [57] recently showing that LCO formation in CrCoNi remains relatively insensitive to cooling rates and annealing treatments. While these alloys differ in crystal structure, they share key common features, including single-phase stability and segregation of a dominant element that does not partake in the preferred LCO motif. This motivates a comparative analysis of FCC and BCC systems with broadly similar segregation behavior and ordering tendencies, enabling identification of transferable mechanistic features that govern near-boundary compositional patterning. Using bicrystal models, hybrid MC/MD simulations reveal simultaneous bulk chemical ordering and grain boundary segregation, with Ni and Nb enriching at the interface in CrCoNi and NbMoTaW, respectively. In CrCoNi, ordering patterns extend over 6 nm into the grain and persist even at high temperatures, indicating a robust thermodynamic driving force from the grain boundary. In NbMoTaW, anisotropic B2-like Mo-Ta ordering emerges at a Σ3 (221)/(001) boundary but weakens as the boundary is rotated. By examining multiple grain boundary orientations and alloys, this study demonstrates that LCO amplification near grain boundaries is a general phenomenon and offers a pathway for future experimental isolation and enhancement of ordering behavior in CCAs.

## II. Computational Methods

An embedded atom method (EAM) potential developed by Li et. al. [15] was primarily used to model atomic interactions in the CrCoNi system. This potential has been previously characterized in-depth [15] and widely applied to study the effect of configurational state on defect evolution such as dislocations [15,28] and grain boundaries [14,58]. Special attention was given to adequately capturing the mixing enthalpies governing LCO formation during potential development. However, discrepancies in chemical correlations relative to DFT calculations [59–61] do exist. For example, while the EAM potential predicts less Ni-Cr ordering compared to DFT, both methods consistently predict Co-Cr ordering and Cr-Cr repulsion. This difference may originate from the lack of explicit spin polarization considerations in the DFT calculations used to fit the EAM potential. To address these discrepancies, a second set of limited simulations was performed with a recently developed neural network potential [62,63] that was designed to more accurately capture the expected chemical ordering behaviors shown in dedicated DFT calculations. Results and discussion from the neural network potential are described in Supplementary Note 2, although the potential was relatively computationally inefficient and there were severe limitations on simulation size and the ability to fully relax the system. Importantly, both potentials predict significant elemental ordering/clustering at grain boundaries, aligning with experimental observations in comparable FCC CCAs [18,19,64]. Atomic interactions for NbMoTaW were modeled using a moment tensor potential that has been extensively used for studying LCO [13,65] as well as its effects on dislocation and grain boundary mechanics [20,31].

Atomistic simulations were performed using the Large-scale Atomic/Molecular Massively Parallel Simulator (LAMMPS) software package [66]. The OVITO [67] software was used for visualization and adaptive Common Neighbor Analysis (CNA) [68] was to identify crystalline structure and defect atoms. For the FCC CrCoNi, a Σ11<110> symmetric tilt grain boundary was chosen due to previous evidence of near-boundary segregation in related alloys [69]. For the BCC NbMoTaW, a series of asymmetric Σ3 tilt boundaries were modeled with varying inclination angles relative to a Σ3(112)<110> ($\varphi = 0°$) symmetric tilt reference. The inclination angles studied were $\varphi = 19.47°$, $\varphi = 27.94°$, and $\varphi = 35.26°$, corresponding to terminating planes of (111)/(1 1 -5), (775)/(1 1 -11), and (221)/(001), respectively. Starting grain boundary configurations were identified using an iterative sampling approach for finding low energy structures at 0 K [70]. Each Σ11 simulation cell contained 135,168 atoms with dimensions at 0 K of $L_x \sim 13$ nm, $L_y \sim 28$ nm, and $L_z \sim 4$ nm and periodic boundary conditions in all directions (Figure S1(a)). Σ3 simulation cell sizes ranged from 47,040 to 58,968 atoms, with aspect ratios similar to those of the Σ11 models. Initial samples were constructed from a single-element lattice followed by random atomic substitution to achieve the desired compositions.

Equilibration protocols were consistent between alloys, with minor differences to accommodate the characteristics of each interatomic potential. In CrCoNi, samples were first annealed at each target temperature for 50 ps in the isothermal-isobaric (NPT) ensemble using a 2.5 fs MD timestep. Hybrid MC/MD simulations followed in the NPT ensemble using the variance-constrained semi-grand canonical ensemble (VC-SGC) [71], which mitigates limitations of traditional canonical and semi-grand canonical ensembles for systems exhibiting phase segregation or precipitation. To maintain equiatomic composition, chemical potential differences were fixed at $\Delta\mu_{Ni\text{-}Co} = 0.021$ eV and $\Delta\mu_{Ni\text{-}Cr} = -0.031$ [15], with the variance parameter κ equal to $10^3$. MC trial swaps were attempted every 20 MD steps for one-quarter of the atoms, using the Metropolis criterion. In NbMoTaW, samples were similarly heated to the target temperature and annealed in the NPT ensemble with an MD timestep of 5 fs. Compositional equilibration was performed by attempting pairwise atomic swaps for 2% of the total atoms every other MD step. Interface concentrations and potential energy were monitored to evaluate chemical and structural convergence, as discussed for CrCoNi in Supplementary Note 2. Finally, all configurations were relaxed via conjugate-gradient energy minimization to eliminate thermal vibrations and enable direct structural comparisons.

## III. Results and Discussion

### A. Near-boundary segregation and LCO amplification in CrCoNi

Ordering and segregation trends in FCC CrCoNi were examined first to establish a foundation for interpreting the near-boundary compositional patterns. Figures 1(a-c) show the evolution of the grain boundary composition during hybrid MC/MD simulations at 400 K, 600 K, and 800 K. All three interfaces begin with equiatomic concentrations (denoted by the black dashed lines) but quickly become enriched in Ni and depleted in Cr and Co. At lower temperatures (e.g., 400 K), Ni concentration initially rises to ~80%, then gradually decreases to ~70%, reflecting a early-stage segregation and the onset of bulk chemical ordering. Wynblatt et al. [72] identified three primary drivers of segregation in multicomponent alloys: elastic strain minimization, chemical interactions, and surface energy. In dilute alloys, segregation is often dominated by strain relief, where larger atoms preferentially occupy interface sites to reduce bond distortion. In CrCoNi, early Ni enrichment aligns with this mechanism since Ni has the largest atomic volume in the disordered bulk (solid shading in Figure 1(d)) and shows a tendency to cluster, making segregation energetically favorable. However, as LCO develops, the energetic landscape changes due to the shift in local atomic environments. Chemical ordering increases the average atomic volumes of Co and Cr (Figure 1(d)), reducing the disparity among species and re-balancing segregation forces. These local structural transformations may be associated with a frustrated strain glass-like transition in Co-Cr clusters [52], which modifies local symmetry, suppresses long-range order, and influences segregation pathways. Once substantial LCO is established, segregation becomes increasingly governed by chemical interactions rather than solely size effects, leading to enhanced Co and Cr concentrations at the interface and a partial redistribution of Ni back into the bulk. The effect is most pronounced at lower temperatures, where chemical interactions are strongest. Because the surface energies of Cr, Co, and Ni are relatively close, with slight differences favoring Ni and Cr, surface energy provides at most a minor tertiary contribution [73]. Overall, this dynamic interplay between strain relief and chemical ordering is consistent with the suppressed surface segregation observed in other strongly ordered alloys [74].

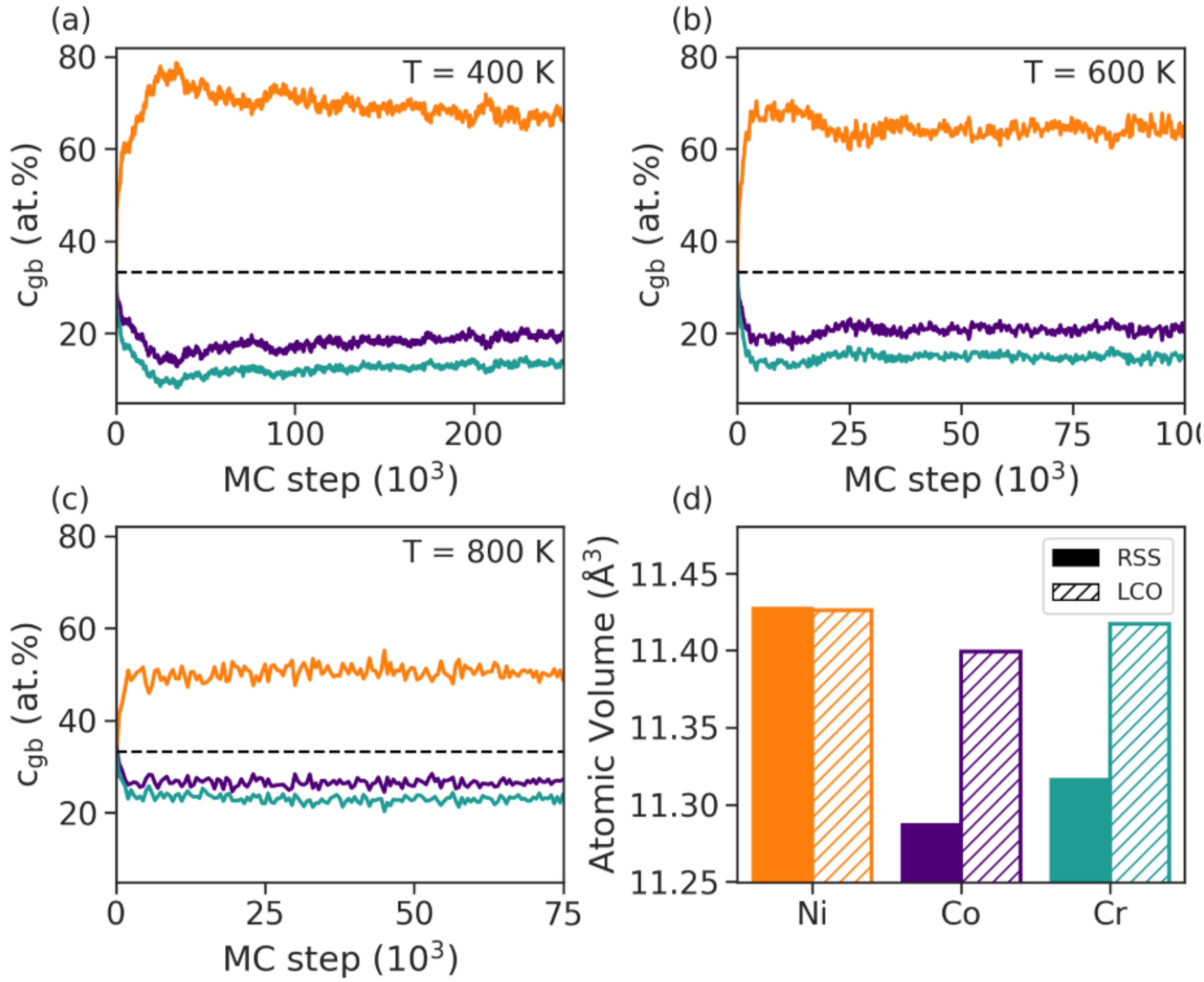

**FIGURE 1. Grain boundary compositions as a function of MC step at (a) 400 K, (b) 600 K, and (c) 800 K, respectively. (d) Average atomic volumes for Ni, Co, and Cr in random solid solution (RSS, solid shading) and in local chemical order (LCO, hashed shading).**

Figure 2(a) shows the equilibrium grain boundary compositions for temperatures from 300 K to 1000 K (homologous temperatures $T/T_m$ of ~0.17 to ~0.59 [75]). The composition is averaged over the final 5000 MC steps to account for minor fluctuations in segregation (see Figure 1), represented by the error bars in Figure 2(a). Ni enrichment remains between 61% and 68% from 300 K to 700 K, then decreases towards equiatomic concentrations at higher temperatures. Co and Cr exhibit similar depletion rates. These trends are consistent with atom probe tomography analysis of FeMnNiCoCr alloys, which show Ni segregation and Cr depletion at low-angle and special high-angle boundaries aged at 450 °C [18]. Cao et al. [14] also predicted similar segregation for a high-angle Σ5 boundary using the same EAM potential, with Ni enrichment exceeding 90 at.% for intermediate temperatures. The disparity in segregation magnitudes between that study and our results here can be explained by density-based thermodynamic models [18], which suggest that low-energy, high-density grain boundary structures (e.g., twin or Σ11) offer weaker trapping environments for solute atoms compared to general high-angle boundaries.

Figure 2(b) shows the arrangement of solute species in defect sites at 300 K, viewed looking into the grain boundary plane. Ni (orange) atoms form distinct clusters, separated from the mixed Co (purple) and Cr (green) atoms. This pattern resembles spinodal-like decomposition seen in FeMnNiCoCr [19] and Fe-Mn [76], and highlights the influence of neighbor interactions and co-segregation over site-specific relaxation. To quantify such ordering in the bulk, the Warren-Cowley order parameter [77] is commonly used [7,15,30]. In its standard form, local atomic concentrations are compared to global concentrations to describe element pair ordering within a defined range. However, this method becomes problematic when applied to grain boundaries, where compositions often deviate significantly from global concentrations due to segregation. To address this, a modified grain boundary order parameter can be defined as:

$$\alpha_{n,GB}^{ij} = 1 - \frac{p_n^{j,i}}{c_{j,GB}}. \quad (1)$$

In Equation (1), $\alpha_{n,GB}^{ij}$ quantifies the interaction between a central grain boundary atom of species *i* and its neighboring atoms of species *j* in the *n*th neighbor shell. In this shell, $p_n^{j,i}$ represents the local concentration of species $j$ around species $i$, while $c_{j,GB}$ is the concentration of species $j$ in both grain boundary and neighboring crystalline sites within the $n$th shell of grain boundary atoms, calibrated for each temperature to account for variations in segregation state. In the absence of preferential ordering, $\alpha_{n,GB}^{ij}$ will be zero. Negative $\alpha_{n}^{ij}$ values indicate strong attractive interactions while positive values indicate repulsive interactions.

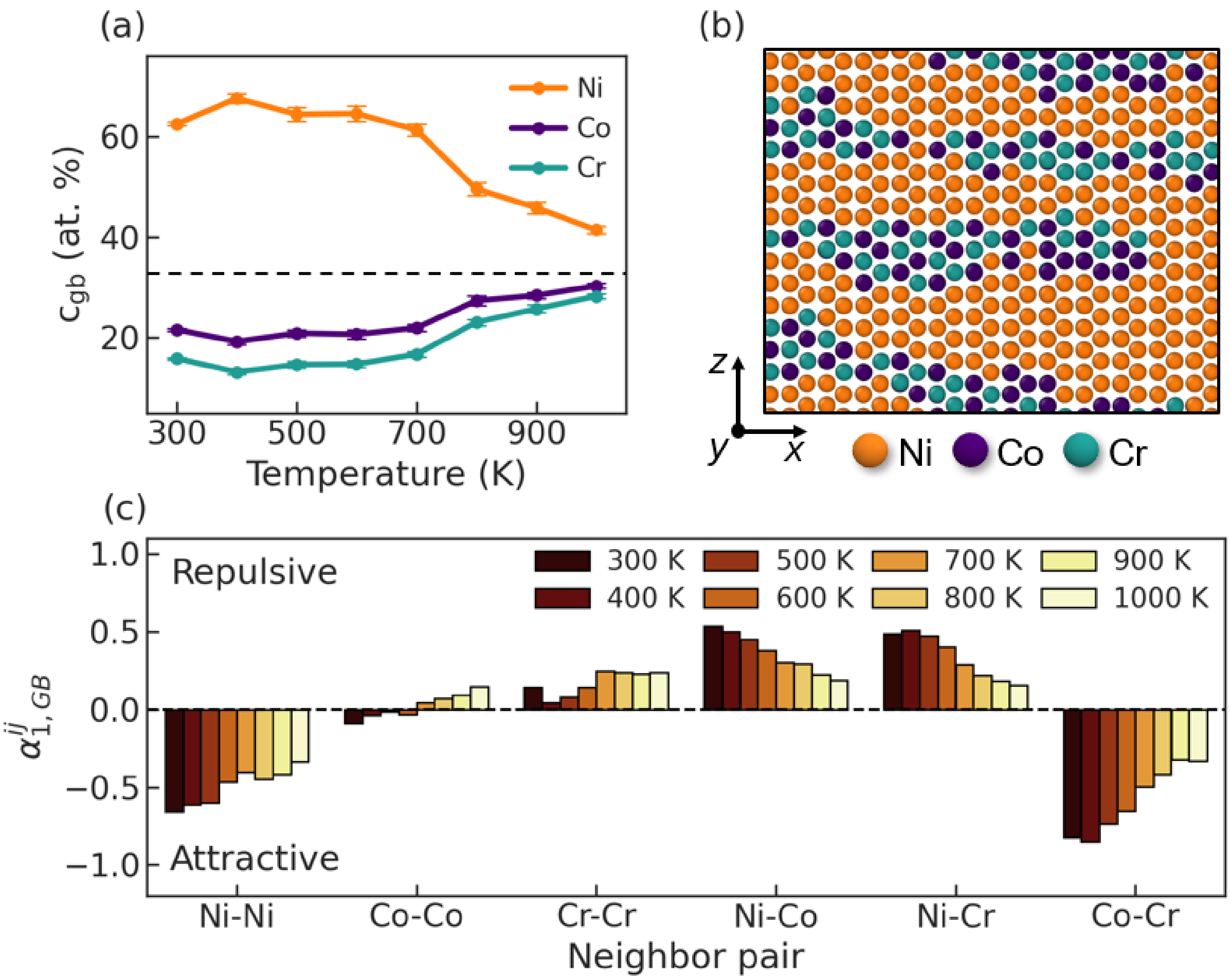

**FIGURE 2. Grain boundary composition and chemical ordering behavior at temperatures from 300 to 1,000 K. (a) Grain boundary concentrations as a function of temperature, with the dashed line indicating equiatomic composition. (b) Elemental configurations in the grain boundary at 300 K, viewed normal to the grain boundary plane. (c) The grain boundary order parameter for pairwise species interactions as a function of temperature.**

Figure 2(c) presents the grain boundary order parameter for each temperature and species pair. Ni-Ni and Co-Cr interactions dominate, as reflected by their large negative values compared to the mostly positive (repulsive) interactions of other pairs. Despite significant Ni segregation (Figure 2(a)), the magnitude of Ni-Ni interactions is smaller than that of Co-Cr, pointing to a weaker clustering effect for Ni and a greater chemical affinity for Co-Cr due to co-segregation and lower concentrations. For example, although Ni grain boundary enrichment exceeds 60 at. % at 400 K, $\alpha_{1,GB}^{Ni-Ni}$ is only -0.62 (~50% increase relative to the bulk), whereas $\alpha_{1,GB}^{Co-Cr}$ reaches -0.85 (~13% decrease relative to the bulk). The less negative Ni-Ni order parameter for grain boundary interactions does not imply that Ni-Ni pairs are uncommon. Instead, it shows that Ni atoms are not preferentially clustering relative to their elevated concentration. Thus, Ni clustering does not scale proportionally with Ni segregation. These trends are consistent with prior observations in single-crystal simulations by Jarlöv et al. [78], where increasing the global Ni concentration was shown to suppress Ni clustering. However, while those authors demonstrated that deviations in composition drive changes in effective chemical interactions at the bulk scale, our results extend this understanding by showing that defects can locally modify the effective interactions even in a closed, composition-fixed system. Grain boundaries therefore act not only as segregation sites but also as regulators of local chemical affinity that can reshape ordering tendencies relative to the bulk.

Previous studies have shown that segregation in CCAs occurs not only at defects but also in the surrounding crystalline regions [13,32,69,72]. Fully characterizing segregation and ordering at interfaces requires analyzing both the grain boundary itself and the adjacent near-boundary zones. To illustrate the compositional variation within the bicrystal after equilibration, concentrations were measured in atomic planes parallel to the grain boundary along the [113] direction. Figure 3(a) provides a spatial reference for plane indexing, while Figures 3(b)-(d) show corresponding elemental concentrations at 300 K. The highest Ni enrichment appears at the grain boundary core, specifically at Planes 0 and 132 in Figure 3(b). Compositional fluctuations are also evident near the grain boundaries and, to a lesser extent, within the bulk. Figure 3(c) zooms in on the boundary region centered at Plane 0 (red box in Figure 3(b)) to reveal the extended segregation profile. Within the boundary core (Planes -1, 0, and 1), Ni reaches its peak concentration, while Cr and Co exhibit alternating peaks, indicating persistent LCO even at the interface. Beyond the boundary, compositional oscillations extend into the grain interior, forming periodic Ni-rich and Ni-depleted waves along the [113] axis, with antiphase fluctuations in Co and Cr. This pattern originates at the boundary and spans ~60 atomic planes (~6 nm). The most prominent Ni-depleted region adjacent to the grain boundary coincides with high Co and Cr concentrations (black arrows in Figure 3(c)). Deeper into the grain, Ni enrichment exceeds 50 at.%, and the oscillation amplitude gradually diminishes, transitioning toward a more uniform bulk composition.

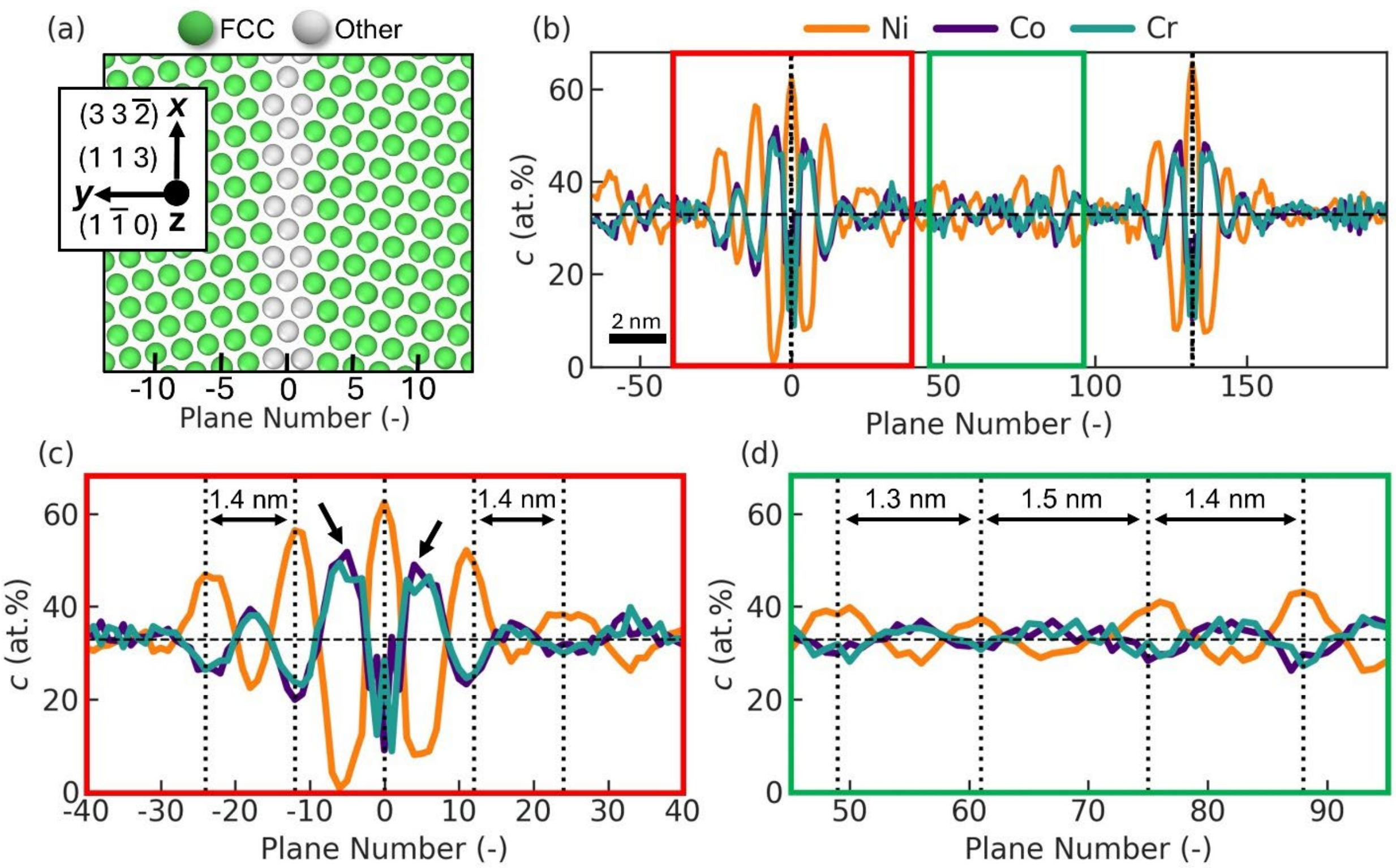

**FIGURE 3. Grain boundary structure and its composition profiles at 300 K. (a) Atomic configurations of the Σ11 grain boundary, with boundary atoms in white and FCC atoms in green. (b) Elemental concentration profile, with dotted vertical lines highlighting the central grain boundary planes. Red and green boxes in (b) indicate the regions analyzed in (c) grain boundary and (d) bulk, respectively. The dotted vertical lines in (c) and (d) are centered on Ni peaks and show the ordering wavelength of ~1.4 nm. Arrows in (c) highlight the most prominent Ni-depletion region and corresponding Co-Cr enrichment.**

Compositional waves were also observed in bulk regions due to LCO formation, although of significantly lower amplitude. Figure 3(d) presents compositional data from a bulk region (green box in Figure 3(b)), showing that oscillations align along [113] even in the absence of a grain boundary. Dotted lines in Figure 3(d) highlight the Ni concentration peaks, with the peak-to-peak distance representing the approximate ordering wavelength, spanning ~14 planes. Although Ni concentrations in the bulk only fluctuate between 27 and 40 at.%, the ordering wavelength is similar to that near the grain boundary region. Both regions demonstrate spatial ordering wavelengths of ~1.4 nm, which can be interpreted as the average ordered domain size, a useful metric for comparison with experimental observations [6,38].

These bulk compositional waves suggest that LCO should be experimentally detectable in this system. However, TEM specimens are typically one to two orders of magnitude thicker than the bicrystal model used here, meaning that chemically ordered regions would be buried among larger volumes of material without the same order as one goes through the thickness. To explore this idea further, a thicker bicrystal model ($L_z$ = 20 nm) was equilibrated at 500 K, and planar compositions were again analyzed. Figures 4(a) and 4(b) compare concentration profiles from the thinner and thicker bicrystals, respectively. In both cases, clear compositional patterning is observed at the grain boundaries (dotted vertical lines), but the amplitude of concentration fluctuations within the grain interior (highlighted by black boxes) is substantially reduced in the thicker sample. This smoothing results from the overlap of Ni-Ni and Co-Cr clusters along the beam direction, a depth-averaging effect that would be even more pronounced in typical TEM specimens with thicknesses of 50-100 nm. Importantly, the compositional patterning at the grain boundary remains unaffected by the increased sample thickness.

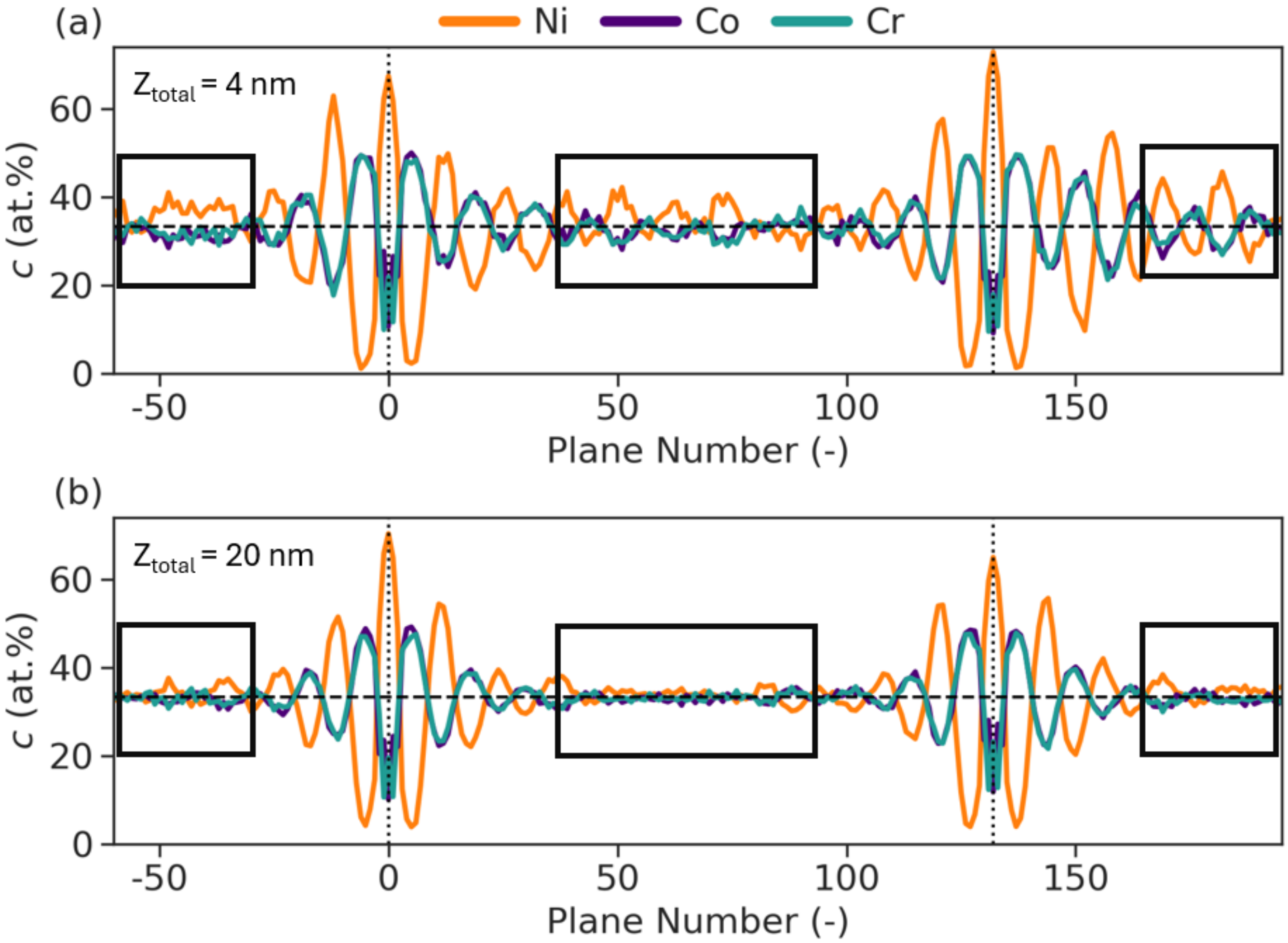

**FIGURE 4. Local element compositions of samples with two different thicknesses at 500 K. (a) Planar compositions for a thinner sample (4 nm). (b) Planar compositions for the thicker sample (20 nm). Black bounding boxes highlight bulk compositional fluctuations in the two samples.**

Figure 5 shows the segregation profiles across a temperature range of 300 K to 900 K, averaged over two grain boundaries per cell. Recall that the grain boundary spans planes -1 to 1 in each plot, where Ni enrichment is most pronounced. While the grain boundary segregation profile remains qualitatively consistent across all temperatures, its magnitude decreases notably at 700 K and 900 K. Compositional patterning extends from the interface in each case, with increased attenuation observed at higher temperatures. For example, only a single Ni-enriched peak near the boundary is visible at 700 and 900 K (black arrows in Figures 5(c) and (d)), whereas multiple peaks are observed at 300 K and 500 K (black arrows in Figures 5(a) and (b)). The broader compositionally-templated zone at 500 K compared to 300 K indicates stronger alignment with bulk chemical ordering perpendicular to the interface. This behavior reflects the energetic interplay between structure and chemistry: at 300 K, favorable bulk chemical ordering leads to larger clusters and isotropic domain growth within the grain, while at 500 K, a balance between anisotropic, structure-driven oscillations and bulk LCO formation results in a wider oscillation profile spanning approximately 7 nm.

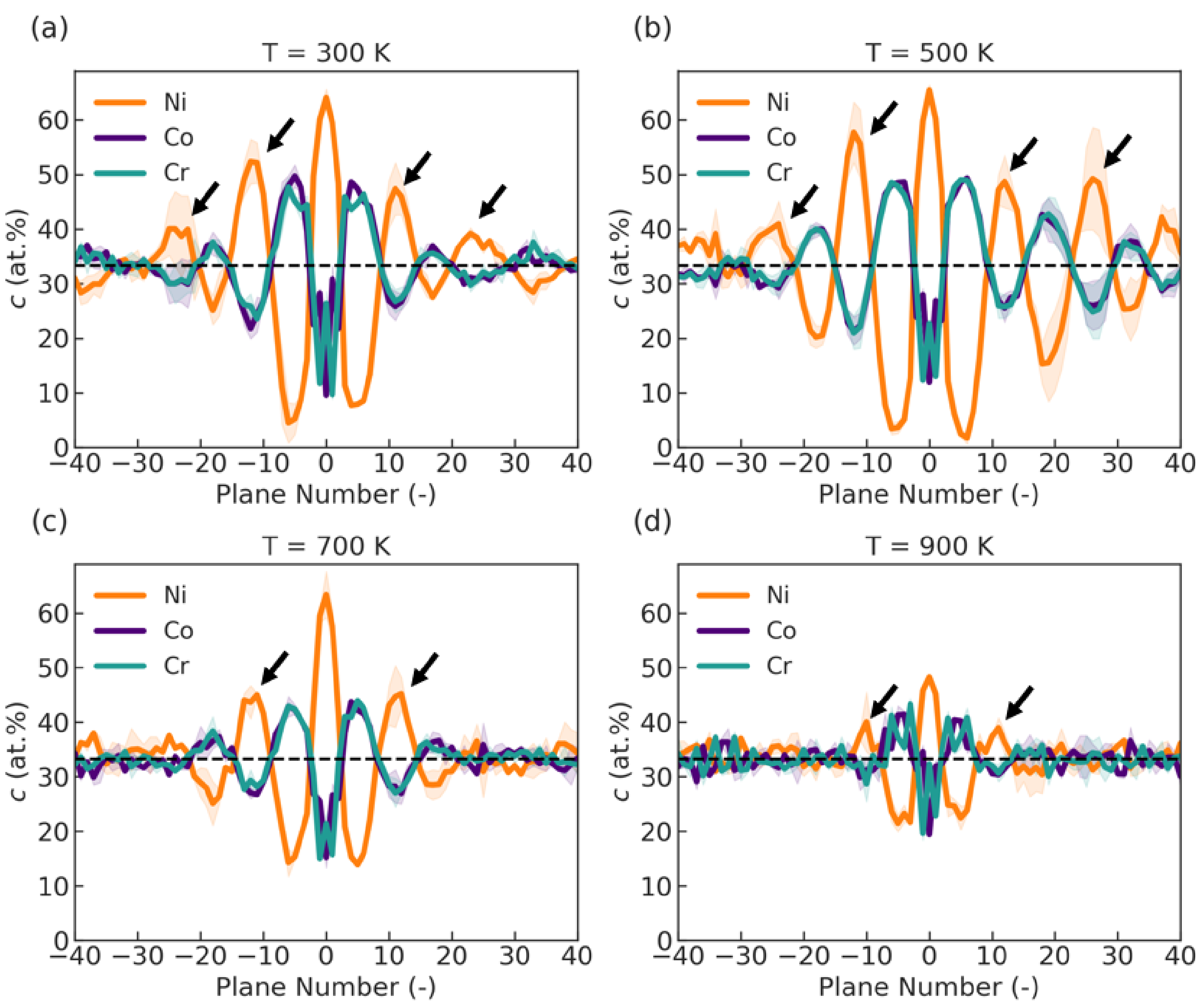

**FIGURE 5. Near-boundary compositional patterning at temperatures of 300, 500, 700, and 900 K. Arrows in each plot highlight the Ni-enriched peaks that diminish in magnitude with increasing temperature.**

Besides the sparse distribution and small size of LCO clusters within grains, experimental detection is further complicated by chemical disorder induced by the high processing temperatures used for homogenization and heat treatment. As temperature increases, the degree of LCO overall decreases due to the growing contribution of entropy to the Gibbs free energy. Although nanobeam electron diffraction patterns have shown extra reflections related to LCO in CrCoNi after annealing at 873 K and 1273 K [5], much of the low-temperature bulk ordering is diminished by 900 K [15].

Our simulations show a similar trend in bulk regions, where LCO diminishes significantly above 700 K. However, even at elevated temperatures (Figure 5(c) and (d)), grain boundaries continue to localize this behavior, albeit less effectively than at lower temperatures. Notably, Co-Cr clustering remains above the bulk average even at 900 K, indicating that chemical disordering is locally delayed near the boundary. At these higher temperatures, the compositional pattern suggests that a structural driving force, likely from elastic distortions in planes parallel to the boundary, persists as the strength of chemical interactions declines. The interplay of these driving forces stabilizes LCO and should enable its characterization after annealing at temperatures sufficient for atomic diffusion.

To investigate how the boundary structure and segregation state influence near-boundary patterns, the local structure at and near the interface was analyzed under two additional model conditions at 300 K: (1) a random solid solution condition and (2) a fully Ni-saturated condition. Figure 6 presents atomic models of these two boundaries along with their planar compositions and atomic volume profiles, with an MC/MD equilibrated boundary included for comparison. In the random solid solution, both the grain boundary and bulk are chemically disordered, while in the Ni-saturated boundary all defect sites are occupied by Ni atoms. Atomic volume was used to assess structural state as well as its correlation with segregation. In the random solid solution sample (Figure 6(a)), minor volumetric changes occur near the interface, with local atomic volumes decreasing by 0.35% for Ni and Co and 0.44% for Cr in Planes 3 and -3, compared to bulk averages (dashed lines in the third row of Figure 6(a)). These planes correspond to the first Co/Cr-enriched plane parallel to the interface in the equilibrated samples. Such structural distortions are intrinsic to the interface structure and have also been predicted in pure metals [69]. In the Ni-saturated boundary, atomic volume decreases by more significant values of 2.0%, 2.3%, and 2.8% for Ni, Co, and Cr, respectively, in Planes 3 and -3 (third row in Figure 6(b)). In CrCoNiFe-based CCAs, the near-boundary sites with maximum elastic compression served as favorable segregation sites for smaller Fe atoms [69]. However, at the Ni-saturated boundary here (Figure 6(b)), the compressed region extends even further into the material away from the grain boundary defects. This extended compression arises from pronounced Ni segregation, which increases the grain boundary free volume and induces greater structural disorder in the adjacent crystalline sites.

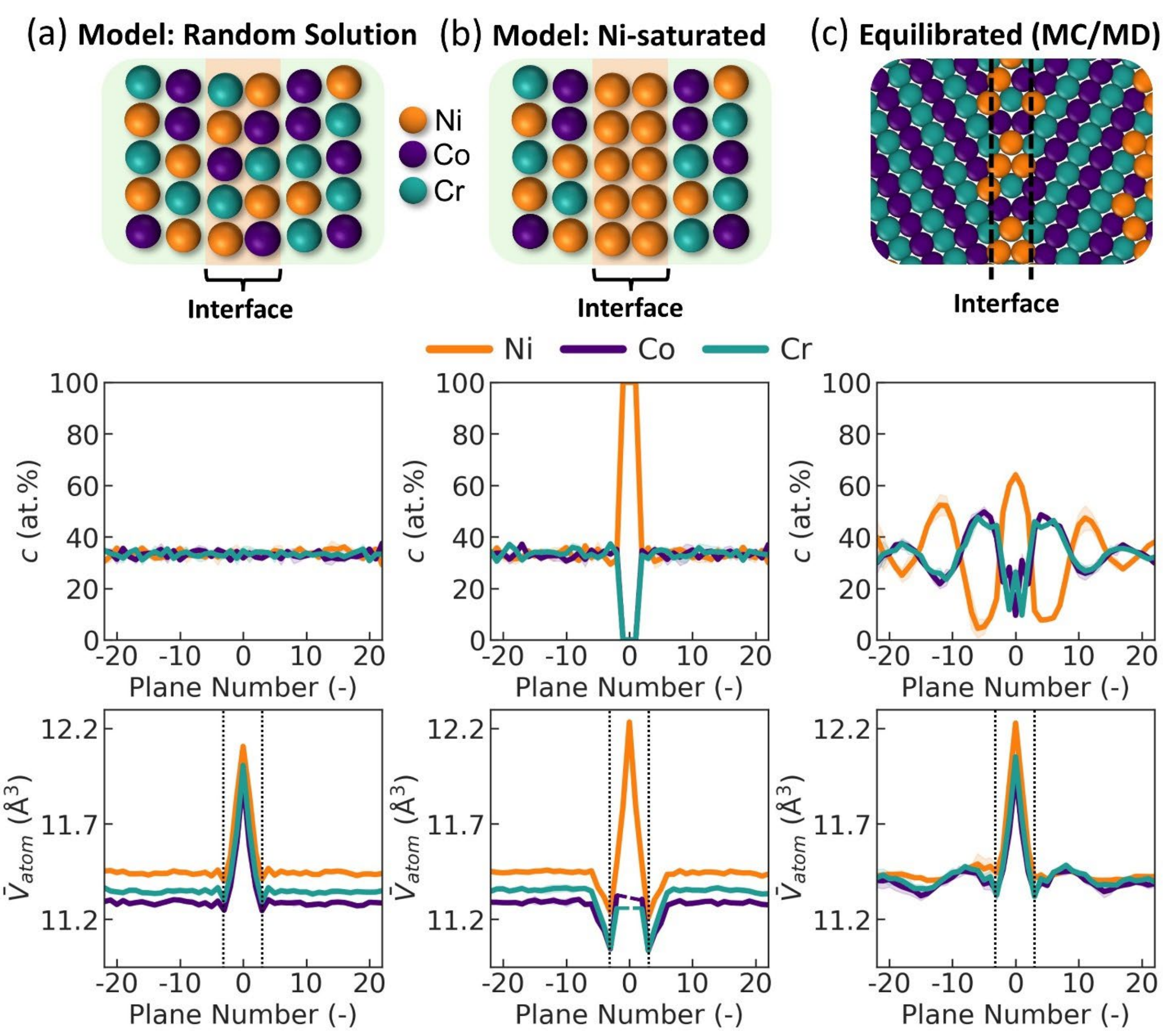

**FIGURE 6. Structural and compositional variations for two limiting grain boundary conditions compared with an equilibrated boundary. Slices parallel to the grain boundary plane compare the random solid solution (a, first column), Ni-saturated boundary (b, second column), and an equilibrated boundary (c, third column). Average atomic volumes for atoms within each plane and ordering condition are shown in the last row, where the dashed lines mark Planes 3 and -3, which exhibit the lowest atomic volume.**

The effect of segregation on the equilibrated interface (Figure 6(c)) is more nuanced because the MC/MD process allows both and chemistry and structure to relax, but the key trends match those seen in the model systems. Ni segregation induces compression in near-boundary planes, which subsequently attracts Co and Cr, mirroring the Co/Cr segregation observed at compressive regions of dislocations [79]. As this localized compression dissipates with increasing distance from the boundary, atomic volumes for all elements expand, creating conditions favorable for a secondary wave of Ni enrichment and giving rise to the near-boundary Ni peak. This structural oscillation, coupled with defect segregation, drives the compositional pattern. At the same time, there are cases where segregation alone at more disordered interfaces can generate related patterns, though to a smaller extent. Additional simulations on a more structurally disordered, high-angle grain boundary show that Ni segregation can still trigger compositional fluctuations, but with much lower amplitudes and smaller effective widths (Supplementary Note 3). These results show that Ni segregation provides a baseline localization effect, but strong amplification requires an interface with a uniform and symmetric structural motif such as the Σ11 boundary.

Further comparison of the ordering patterns in the grain interior and near the boundary clearly shows that atomic configurations are consistent across both regions. This similarity, along with the comparable characteristic ordering wavelength shown in Figure 3, suggests that the underlying chemical interactions driving LCO formation are fundamentally the same in both environments. However, the local distortion and segregation at the interface spatially localize the LCO waves, offering a unique opportunity to study intrinsic LCO structures, unlike the random, diffuse ordering observed throughout the grain.

**B. LCO amplification in non-equiatomic NbMoTaW**

To demonstrate that segregation-driven LCO amplification and localization are not exclusive to a specific material, other alloys should be explored. NbMoTaW is a refractory BCC alloy that exhibits both strong segregation tendencies and pronounced chemical ordering behavior. Although NbMoTaW has been widely studied for its mechanical properties [32,80,81], it has also served as the basis for numerous predictions of LCO, particularly involving Mo-Ta ordering in a B2-type structure [31,32,56,82]. In addition to the dominant Mo-Ta interaction, secondary, less pronounced Nb-Mo and Nb-W interactions are also predicted in the bulk. These preferences are likely caused by differences in atomic size and electronegativities between Group 5 (Ta, Nb) and Group 6 (Mo, W) elements [83], with the Mo-Nb interactions being strongest in the regions where Ta is depleted, such as near grain boundaries [65]. A key similarity to the CrCoNi system is that the primary segregant (Nb in this case), does not directly participate in the dominant B2 structure, which instead involves Mo and Ta. Here, segregation and grain boundary-enhanced LCO in NbMoTaW is examined to further generalize the behavior observed for CrCoNi.

The B2 structure is defined by two interpenetrating sublattices with distinct elemental occupancies, which, in long-range ordered systems, give rise to alternating {100} atomic planes enriched in different species. An example of B2 (Mo,Ta) LCO within the grain of a Nb10-Mo25-Ta25-W40 alloy equilibrated at 300 K is shown in the black boxes in Figure 7(a). This composition is selected to mitigate the strong tendency for Nb clustering near grain boundaries in equiatomic NbMoTaW alloys, as will be discussed later. Comparison of the Warren-Cowley order parameter in bulk sites between the two compositions (Figure 7(c)) shows qualitatively similar ordering behavior, with minor variations arising from the change in global composition. For example, Mo-Ta remains the preferential chemical interaction while Nb-Nb affinity decreases in the non-equiatomic alloy. This reduced affinity reflects the shift in energetic preference away from self-clustering toward increased mixing, but the overall trends indicate the same LCO structures dominate for the non-equiatomic composition. In Figure 7(a), the observed planar ordering preferentially forms along [001] directions and is characterized by alternating planes of unlike species. Accordingly, a boundary terminating on a {100} plane serves as a natural template for studying the effect of grain boundaries on ordering tendencies, while an asymmetric orientation allows for the role of lattice misorientation to be assessed.

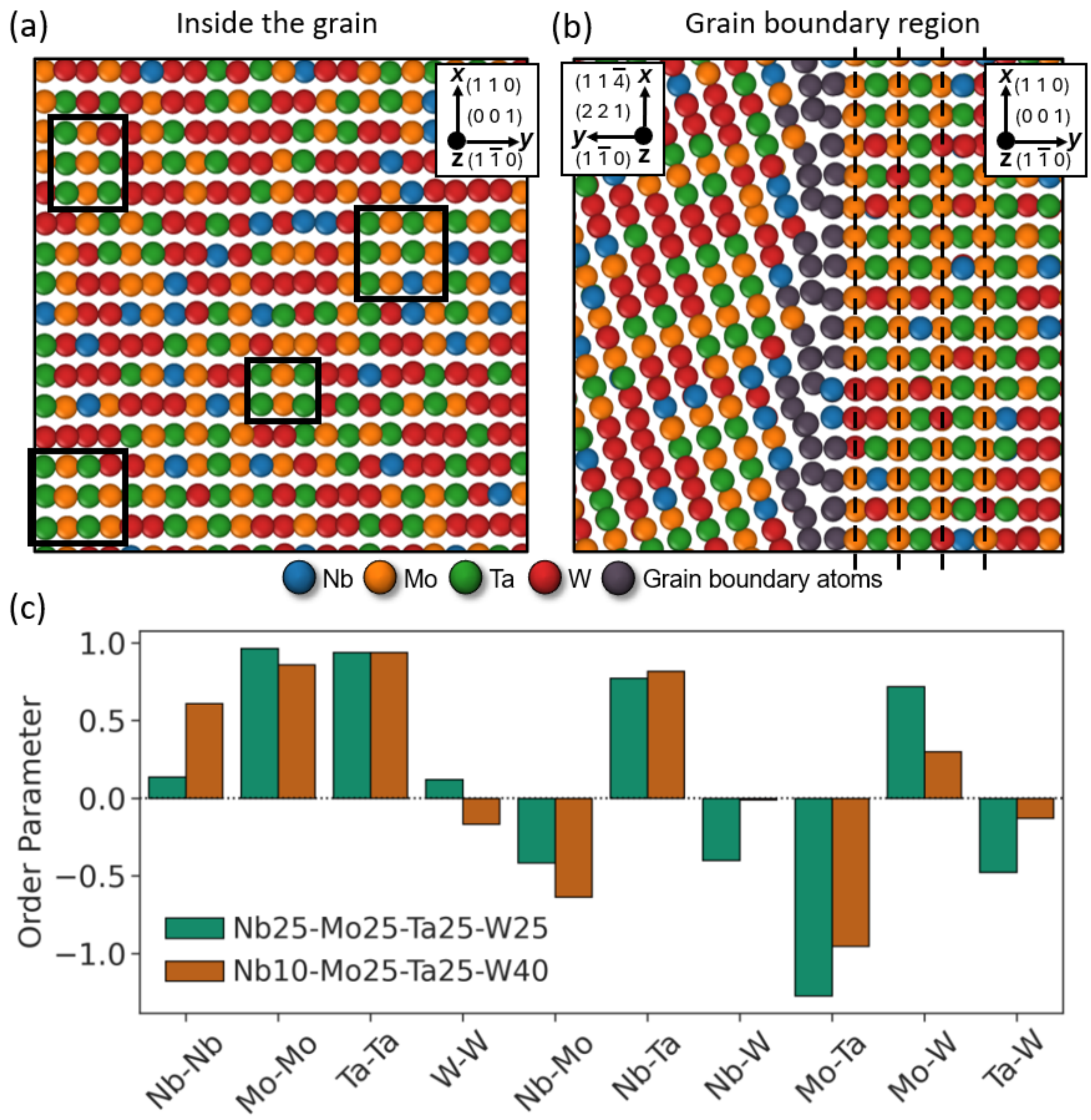

**FIGURE 7. Atomic configurations within the grain and near the grain boundary and quantitative ordering analysis in bulk sites. (a) A region inside the crystalline part, with black boxes highlighting domains of of B2-type LCO. (b) Atomic configurations at an asymmetric Σ3 (221)/(001) boundary. Dashed lines indicate the Mo-enriched planes, while the intervening planes are enriched in Ta. (c) Warren-Cowley order parameter comparison between equiatomic (green) and non-equiatomic (orange) compositions.**

Figure 7(b) presents an asymmetric Σ3 boundary with terminal planes of (221)/(001) that is rotated $\varphi = 35.26°$ from the coherent twin orientation ($\varphi = 0°$). Within the grain (Figure 7(a)), LCO remains sparse and short-ranged as shown by the black boxes. In contrast, on the (001) side of the boundary in Figure 7(b) (right-hand side grain), extended B2-like ordering is apparent. For example, the first BCC plane just to the right of the boundary (dashed line) is strongly enriched in Mo (orange atoms), with the adjacent plane enriched in Ta (green atoms). On this side of the interface, the local boundary structure is atomically flat, allowing for uniform Mo enrichment along the entire boundary plane. This uniformity, in turn, promotes enrichment of a secondary element (Ta) in the adjacent atomic plane, facilitating a layered ordering pattern characteristic of B2-like structures. This alternating pattern persists across ~7-8 planes extending into the grain, while no coherent, medium-ranged LCO is seen on the left-hand side of the boundary. Similar to CrCoNi, while LCO in the bulk remains sparse and intermittent in NbMoTaW, near the boundary it becomes highly concentrated and amplified in magnitude.

The influence of grain boundary structure becomes even more evident when the Σ3 boundary is incrementally rotated, producing slight variations in inclination angle and terminal boundary planes. The top rows of Figures 8(a-c) present three asymmetric Σ3 boundary structures with inclination angles of 35.26°, 27.94° , and 19.47°, respectively. The red dashed lines indicate the first atomic row adjacent to the grain boundary atoms. While this plane is flat at 35.26°, it becomes increasingly tilted at lower inclination angles, disrupting the planarity required for consistent elemental enrichment. This structural frustration leads to competition between adjacent planes for elemental occupancy, weakening the coherent ordering sequence. The impact of this disruption is shown in the corresponding planar composition profiles (Figures 8(d-f), bottom row), where the grain boundary region (Planes -1, 0, and 1) is marked gray. For $\varphi = 35.36°$ (Figure 8(d)), Mo and Ta reach peak concentrations of nearly 60 at.% in the first two planes adjacent to the boundary (35 at.% above their nominal values). Although enrichment attenuates with distance from the interface, the compositional patterning remains detectable over 6 atomic planes. For $\varphi$ = 27.94° (Figure 8(e)), a similar trend is observed, but peak concentrations fall to ~45 at.%. By $\varphi$ = 19.47° (Figure 8(f)), the enrichment is confined to only the first two atomic planes and is significantly diminished. Concurrently, Nb segregation at the boundary also decreases with decreasing inclination angle, while Mo segregation increases. As Nb enrichment provides a thermodynamic driving force for Mo segregation via favorable Nb-Mo ordering in the relative absence of Ta [34,82], its reduction further suppresses the development of Mo co-segregation and thus extends B2-like patterns. The increase in Mo grain boundary enrichment also leads to less near-boundary Mo enrichment due to unfavorable Mo-Mo bonding. As both segregation strength and boundary structural compatibility decline, the near-boundary ordering becomes increasingly weak and quickly falls off in magnitude.

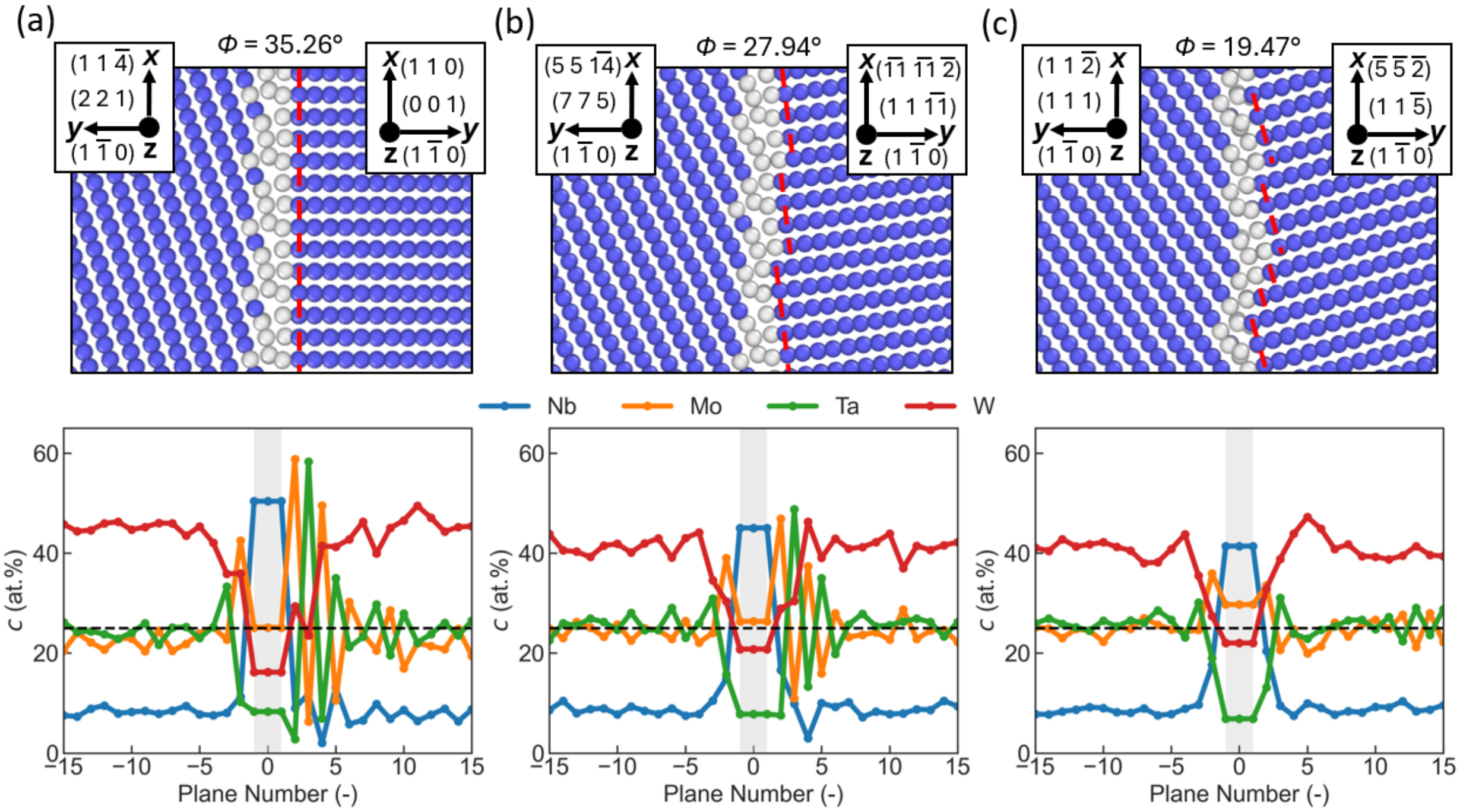

**FIGURE 8. Compositional patterning as a function of Σ3 inclination angle, $\varphi$. Atomic configurations (top row) and corresponding planar compositions (bottom row) near the boundary are shown for (a) $\varphi$ = 35.26°, (b) $\varphi$ = 27.94°, (c) $\varphi$ = 19.47°. In the atomic snapshots, red dashed lines indicate the first atomic plane adjacent to the grain boundary. In the composition profiles, gray shading highlights the grain boundary region (Planes -1, 0, and 1), with concentrations averaged across these three planes to represent interfacial segregation.**

One of the key distinctions between the BCC and FCC alloys studied here lies in the nature of the driving forces governing near-boundary segregation. In the FCC system, structural distortion near the grain boundary plays an active role in localizing chemical order, as shown by clear local drops in atomic volume that stabilize Co and Cr enrichment in adjacent crystalline planes. This structural contribution acts jointly with strong intrinsic LCO to promote extended near-boundary compositional modulation. In contrast, comparable structural distortions are not observed for the BCC NbMoTaW alloy. Figure 9 shows the average atomic volume for each element across a Σ3(221)/(001) boundary in NbMoTaW for both a random solid solution and an equilibrated configuration. In the random solid solution state (Figure 9(a)), the elevated atomic volumes in the grain boundary core relax rapidly to near-bulk values within the first adjacent crystalline planes. After chemical and structural equilibration (Figure 9(b)), the atomic volume profiles become asymmetric due to the evolved segregation state, with the largest changes occurring on the (001)-terminated side of the boundary. However, these variations do not reveal a systemic or oscillatory strain field extending into the grain that could serve as a structural scaffold for localizing LCO. As a result, the dominant driving force for near-boundary segregation in NbMoTaW is chemical in nature, originating from the alloy's intrinsic ordering tendencies rather than by a coupled structural distortion field. The contrasting segregation morphologies and spatial extents observed in FCC and BCC alloys therefore reflect differences in the relative weighting of chemical and structural contributions.

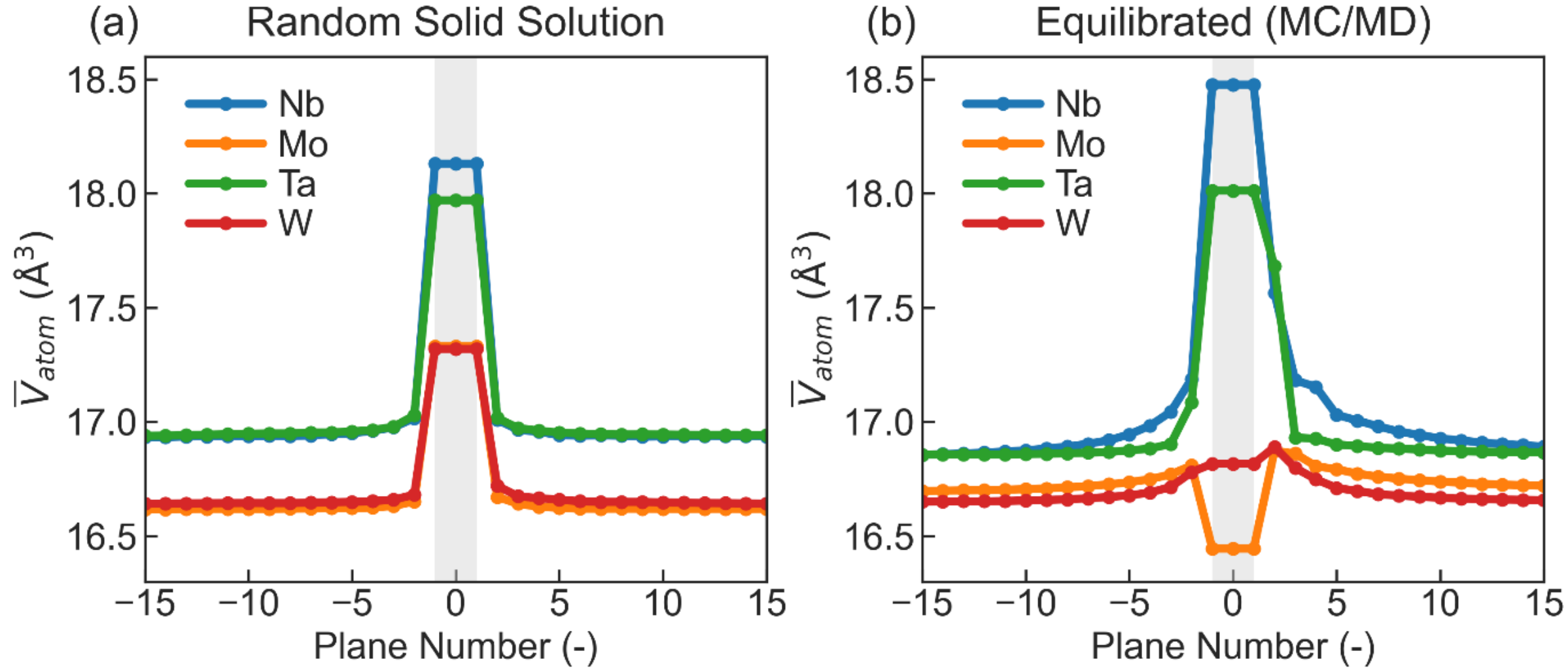

**FIGURE 9. Structural variations in NbMoTaW across a Σ3(221)/(001) boundary. Slices parallel to the grain boundary plane show the average atomic volume for each element in (a) a random solid solution and (b) the equilibrated configuration. Gray shading denotes the grain boundary region, with atomic volumes averaged over the three interfacial planes.**

Building on this mechanistic distinction, the results for CrCoNi and NbMoTaW also demonstrate how alloy chemistry can be deliberately leveraged to control the expression of near-boundary LCO. For example, informed understanding of Nb segregation tendencies, Mo-Ta ordering preferences, and Mo-Nb interactions in Ta-depleted environments (such as at grain boundaries) guided the selection of a modified NbMoTaW composition designed to enhance the visibility of near-boundary ordering. An example of the resulting planar compositions for the asymmetric Σ3(221)/(001) boundary in an equiatomic NbMoTaW alloy is shown in Figure 10. For this composition, Nb segregation is so strong that it dominates the overall response. The plane directly adjacent to the boundary on the (001) side is heavily enriched in Nb due to the tendency for Nb-Nb clustering and the near-boundary Mo-Ta ordering pattern, while present in modest amounts, is significantly reduced compared to the examples shown in Figure 8. By reducing the overall concentration, Nb can still preferentially segregate to the grain boundary yet does not overwhelm the adjacent near-boundary region, as previously reported by Aksoy et al [13]. The small amount of Nb that remains in the bulk interacts minimally with the other atomic species.

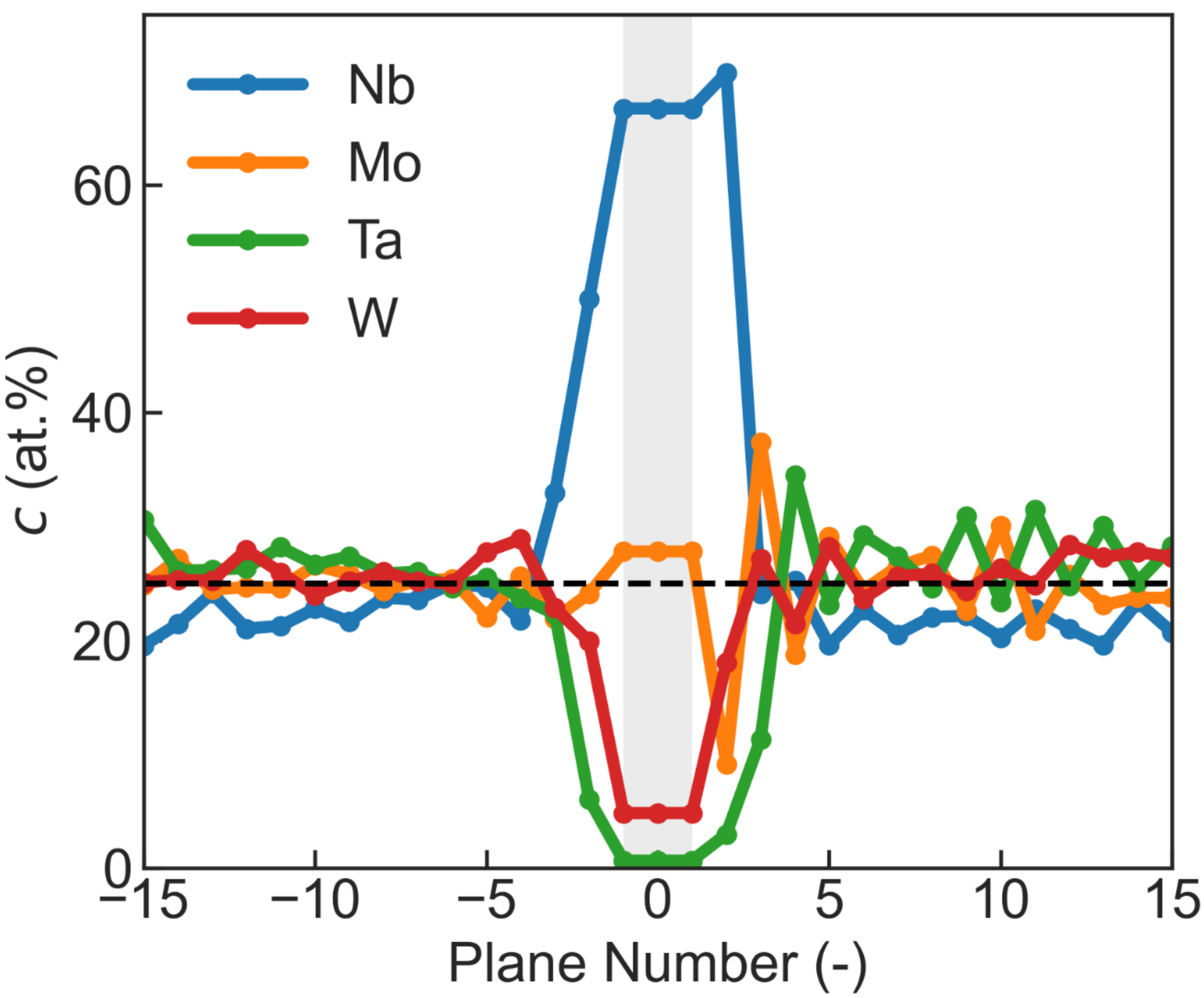

**FIGURE 10. Compositional profile for an equiatomic NbMoTaW alloy at a Σ3(221)/(001) boundary.**

To further demonstrate the combined importance of segregation behavior, crystallographic orientation, and intrinsic ordering tendencies, we performed additional simulations of Nb10Mo25Ta25W40 at a symmetric Σ27(552)[1-10] boundary, as discussed in Supplementary Note 4. These simulations show that Nb and Mo segregate strongly to the Σ27 boundary, consistent with the segregation observed at the Σ3 interface. However, although Mo enrichment extends into the plane directly adjacent to the boundary, no further amplification of LCO develops. This indicates that while Mo enrichment near the boundary is supported by strong Nb segregation, the absence of a suitable terminating plane prevents the emergence of extended ordering motifs. This comparison shows that segregation and interfacial structural distortions are not sufficient to induce LCO in the BCC system. Consequently, experimental efforts aimed at promoting LCO should prioritize identifying grain boundary geometries that provide the crystallographic conditions required for ordering to evolve, rather than assuming broad transferability across boundary types. However, the physical principles developed here suggest that other boundaries may also support LCO amplification when they possess the necessary terminating planes. For example, asymmetric Σ5(430)/(100) and Σ9(447)/(001) interfaces may favor LCO development, as both contain (001) terminating planes on one side. Thus, the Σ3 boundary studied here serves as a model system for identifying the tunable grain boundary features that promote regions enriched in LCO at predictable sites within the microstructure.

Related interface-driven compositional patterning has also been reported in other material systems, including anisotropic compositional domain growth at defects such as surface-directed spinodal decomposition (SDSD) [84]. Jones et al. [85] observed compositional waves with wave vectors normal to a surface in phase-separating poly(ethylenepropylene) and perdeuterated poly(ethlyenepropylene) mixtures, which these authors attributed to the preferential wetting of the latter at the surface. Numerical simulations by Tang et al. [49] further elucidated the SDSD kinetic pathway in crystalline solids, showing that compositional patterning begins with segregation and preferential phase separation at the surface due to misfit stress relaxation, followed by isotropic domain growth within the bulk over time. Subsequent studies demonstrated that interfaces could also induce concentration waves propagating within a sample [48,51,86,87], creating compositional patterns closely resembling those for CrCoNi and NbMoTaW shown above.

In CCAs, analogous mechanisms are likely active due to their diverse decomposition pathways that are inevitably influenced by the presence of defects [88]. However, despite parallels to SDSD-like behavior, CrCoNi exhibits exceptional phase stability in experimental studies [23,55,57,89]. For example, aging treatments performed by Zhang et al. [89] over a temperature range of 600-1000 °C for up to 240 hours yielded single-phase microstructures. While numerous reports indicate the presence of nanoscale ordering motifs, including $L1_0$ and $L1_2$-type structures, these configurations do not evolve into distinct secondary phases. For example, Li et al. [46] and Inoue et al. [90] proposed the formation of $(Ni,Co)_3Cr$ clusters, which may reflect frustrated precursors to phase separation, yet no significant domain growth or transformation was observed.

A key distinction between the hybrid MC/MD simulations performed here and the time-resolved phase-field modeling typically used to study SDSD lies in the treatment of time. Spinodal decomposition is inherently a kinetic process, where time-dependent concentration waves precede long-range phase separation. In contrast, MC/MD simulations sample low-energy configurations by exploring phase space rather than explicitly evolving in real time. In our simulations, this leads to atomic reconfiguration near grain boundaries, where LCO becomes stabilized. As in experimental observations, this may reflect an underlying tendency toward local separation that is energetically or structurally constrained. Li et al. [19] demonstrated that local spinodal enrichments along grain boundary planes in FeMnNiCoCr can act as compositional precursor states to more stable equilibrium phases. This finding highlights how defect chemistry in CCAs may offer fundamental insight into the thermodynamic pathways and the nature of equilibrium states. The tendency for LCO is apparently more active near grain boundaries, where enhanced diffusion kinetics have also been shown to produce depletion zones in certain alloys [91], potentially biasing the nucleation and growth of LCO in adjacent regions. Nonetheless, this emergent nanostructure patterning could be strategically exploited and controlled, underscoring the importance of further investigation into the crystalline regions nearby, but not exactly at, grain boundaries in CCAs.

### C. Comparative analysis of governing principles and alloy-specific expressions

To further contextualize the FCC and BCC results presented above, we next analyze the common physical picture underlying the observed near-boundary compositional patterns and clarify how alloy chemistry and boundary structure jointly govern its characteristics. Figure 11 provides a schematic comparison of the compositional wave behavior using representative planar concentration profiles for Co in CrCoNi and Mo in NbMoTaW, which highlight the dominant enrichment and depletion response associated with the preferred LCO motif in each alloy.

In both systems, the compositional modulation propagates normal to the grain boundary plane and decays with penetration into the grain interior. This shared wave vector orientation indicates interface-directed ordering rather than randomly oriented bulk domains. The modulation amplitude, $A$, can be defined as the difference between the first near-boundary peak and the nominal bulk concentration of the measured element. The decay length, $\xi$, is approximated as the distance from the interface to the point at which the chemical modulation returns to bulk behavior. These descriptors, in addition to the characteristic wavelength, vary between the two alloys because the relative contributions of chemical and structural driving forces differ. In CrCoNi, the long-range chemical modulation extends $\xi \sim 3$ nm into each grain, corresponding to a total patterned zone of roughly 6 nm across the interface. In NbMoTaW, the modulation is much more localized with $\xi \sim 1$ nm. In contrast, NbMoTaW exhibits stronger near-boundary enrichment, with the dominant enrichment layer exceeding the nominal composition by ~33 at.% and producing a high-contrast interfacial signal relative to the sparse and spatially intermittent bulk LCO. Differences in wavelength and motif reflect the intrinsic ordering tendencies of each alloy. In CrCoNi, the modulation consists of coupled Co and Cr enrichment and depletion that occurs in antiphase with Ni enrichment and depletion. In NbMoTaW, the modulation manifests as plane-by-plane B2-like layering of Mo and Ta adjacent to compatible terminating planes.

Although the microscopic expression differs between alloys, the governing factors within the present dataset provide a consistent mechanistic picture. Amplification is controlled by three coupled contributions. (i) The segregation state modifies the local chemistry at the boundary and biases near-boundary pair interactions. (ii) The boundary structure and terminating plane determine crystallographic compatibility with the preferred ordering motif and therefore whether ordering can propagate coherently away from the interface. (iii) The intrinsic LCO tendency of the alloy governs the baseline ordering motif and influences both the wavelength and decay length of the response. In CrCoNi, stronger intrinsic ordering combined with a pronounced atomic volume gradient produces a longer-range response when Ni enriches at the interface and drives secondary redistribution of Co and Cr into adjacent planes. In NbMoTaW, the absence of a pronounced structural distortion field and slightly reduced ordering tendency produce a shorter-ranged effect that is sensitive to boundary geometry. Even when $\xi$ is short, coherent ordering along the interface likely provides a high-contrast, more detectable signal than scattered bulk LCO.

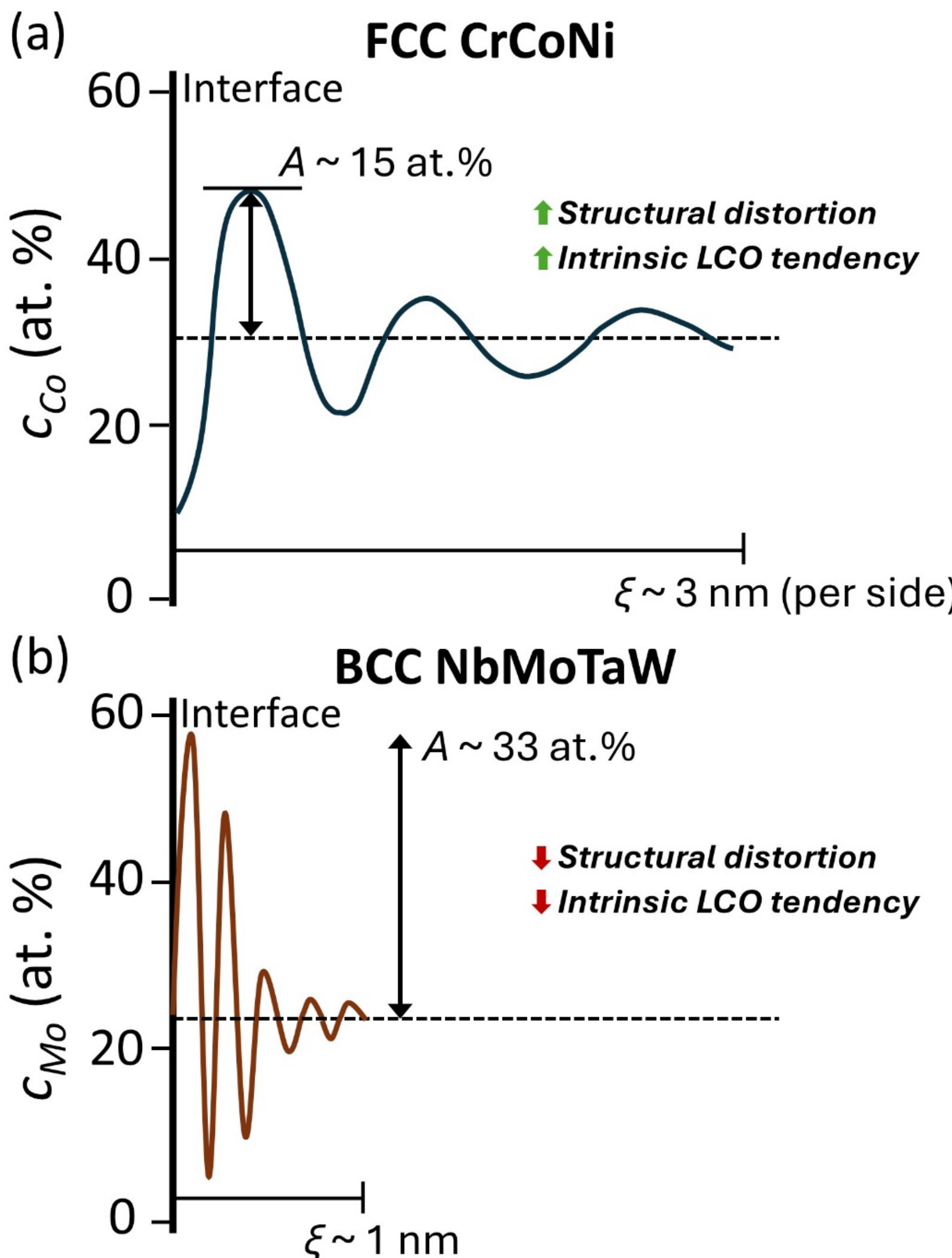

**FIGURE 11. Schematic of near-boundary compositional modulation in (a) FCC CrCoNi and (b) BCC NbMoTaW. Planar concentrations of Co in CrCoNi and Mo in NbMoTaW are shown as a function of distance from the interfacial plane. The modulation amplitude, *A*, is defined as the difference between the first near-boundary peak and the nominal bulk concentration of the measured element. The decay length, *ξ*, is approximated as the distance from the interface to the point at which the bulk behavior dominates. Qualitative descriptors summarize the relative contributions of structural distortion and intrinsic LCO tendency, with green (up) and red (down) arrows denoting stronger and weaker relative contributions, respectively.**

**D. Implications of LCO amplification and localization for experimental investigations**

The predictions of amplified LCO near grain boundaries in multiple CCAs, with different crystal structures, suggest that future experimental efforts should prioritize these interfacial regions within the microstructure when seeking to detect and characterize LCO. To illustrate the significance of the grain boundary patterning for experimental analysis, atomic snapshots from CrCoNi samples were taken at varying depths along the *z*-direction from both grain interior (i.e., bulk) and grain boundary regions within a thicker sample (20 nm depth, as shown in Figure 4) and are shown in Figure 12(a). In each snapshot, the *x* and *y* coordinates were held constant while the depth was incremented by at least 2 nm, exceeding the characteristic ordering wavelength discussed earlier. Figures 12(b) and (c) compare atomic configurations in the bulk and near the grain boundary, respectively. Grain boundary atoms in Figure 12(c) are colored gray to draw attention to the adjacent crystalline sites. In the bulk (Figure 12(b)), Co-Cr ordering and Ni clustering shift positions as depth increases from $Z_1$ to $Z_4$, constrained only by the formation of nearby clusters. For example, the green box highlights a region where a large Ni cluster at $Z_1$ is later replaced by Co and Cr atoms at $Z_2$, with further variations observed at greater depth. This variability complicates LCO measurement, as electrons transmitted through the bulk sample interact with multiple, misaligned cluster configurations. In contrast, Figure 12(c) reveals a more consistent arrangement near the grain boundary. Co and Cr consistently occupy lattice sites across all sampled depths, while dashed green lines mark a persistent Ni-rich layer parallel to the grain boundary plane. This uniform chemical patterning produces a dense, ordered region which should allow for more accurate detection and quantification of LCO, offering a clear target for techniques such as electron diffraction or spectroscopy.

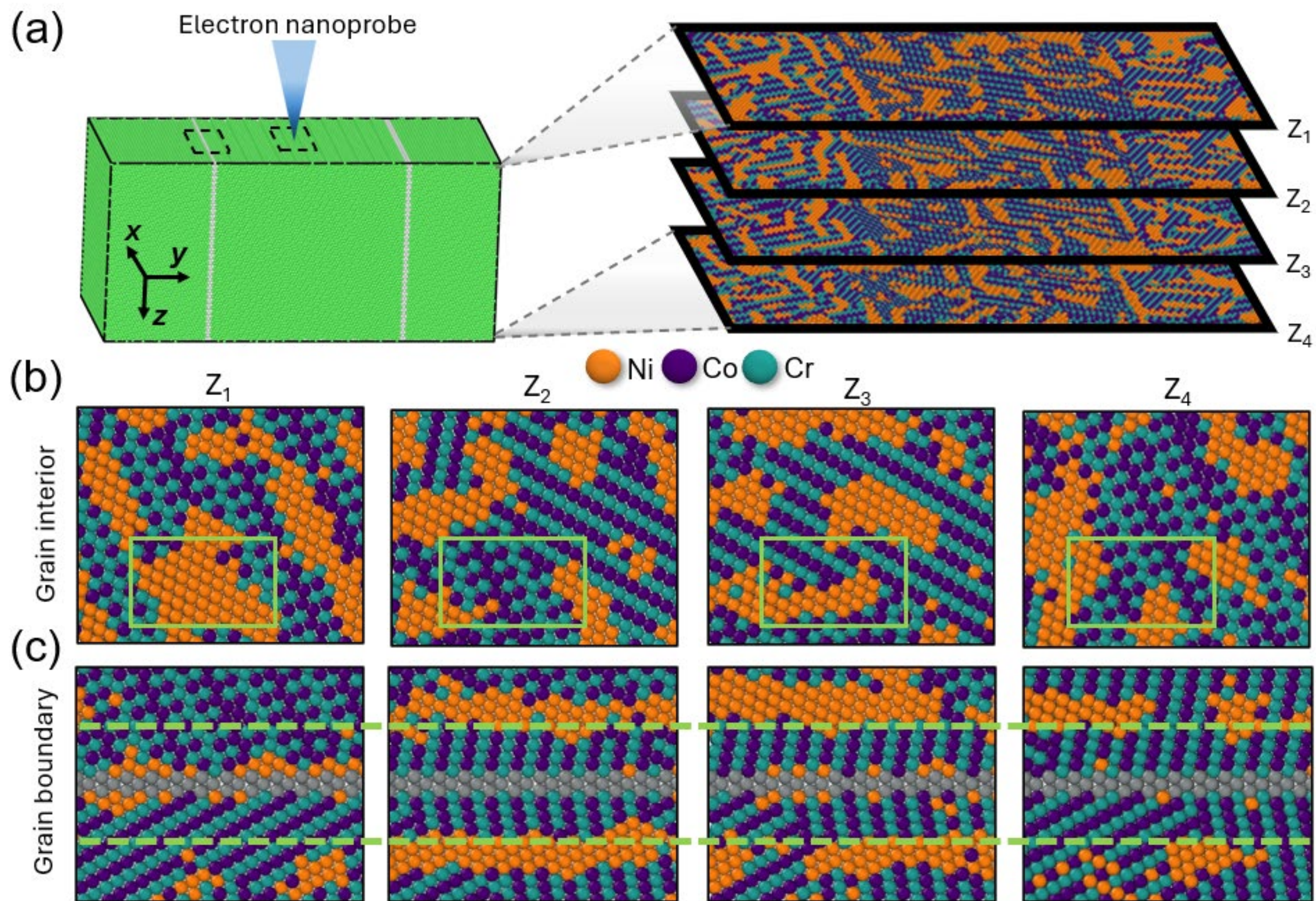

**FIGURE 12. Spatial distributions of LCO throughout the sample. (a) Simulation cell of the thicker sample, colored by structural type, with dashed rectangles indicating approximate coordinates for atomic snapshots taken at varying depths (*z*-direction). (b) Atomic configurations from bulk slices, with depth increasing along $Z_i$. The green box in (b) highlights significant elemental fluctuation through the sample thickness. (c) Atomic configurations from slices near the grain boundary, with boundary atoms colored in gray to emphasize near-boundary composition. Green dashed lines in (c) denote the first layer of Ni enrichment near the boundary.**

The advantages of investigating LCO near grain boundaries versus the bulk are illustrated schematically in Figure 13, where an incident electron beam is aligned parallel to the grain boundary (*x*-direction in Figure 13(a)). At the interface, structural disorder and interfacial segregation act as a template for localizing LCO (Figure 13(b) and (c)). The application of an edge-on electron beam in this region will produce signals rich in structural and chemical information about LCO in the alloy. Furthermore, the LCO orientation is influenced by the grain boundary geometry, offering an additional microstructural lever for optimizing characterization. TEM-based techniques such as nanobeam diffraction or extended electron energy-loss fine structure analysis [92], which employ nanometer-scale probe sizes to resolve local atomic structures, would be particularly well-suited for extracting chemical ordering information from these regions. In contrast, within the grain interior, electrons interact with tiny, isolated LCO pockets that likely vary in orientation and/or can take on different morphologies depending on the selected zone axis (Figure 13(d) and (e)). Any signal from these small regions is watered down by a lack of signal from the surrounding material. Bulk regions sampled at random will therefore yield data that obscures the true LCO, complicating interpretation.

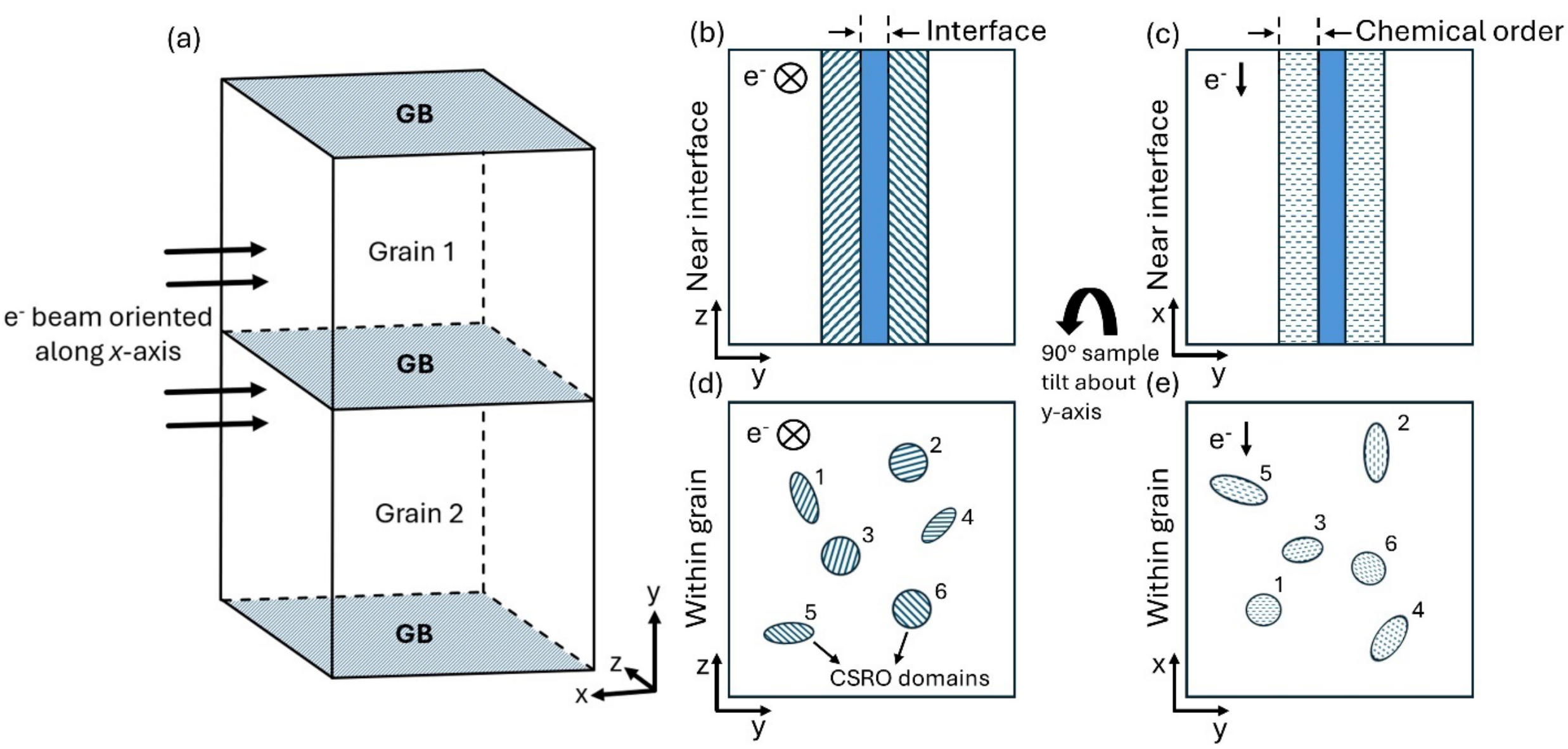

**FIGURE 13. Schematic representation of electron beam interactions with LCO. (a) Orientation and electron beam direction with grain boundary regions abbreviated by GB. (b) *y*-*z* plane view of the interface and near-boundary compositional patterning, illustrating chemical order saturation along the grain boundary plane. (c) Rotation of the sample from (b) about the *y*-axis, showing the identical saturation of chemical order in the *x*-*y* plane. (d) *y*-*z* plane view of LCO within the grain interior, with circular and elliptical shapes depicting scattered**

**regions of localized chemical order. (e) A 90° rotation of (d) around the *y*-axis. In (d) and (e), numbered LCO domains highlight variations in shape and position depending on the zone axis. In (b) and (d), the beam direction is into the page, while in (c) and (e) the beam is oriented top-down.**

The spatially localized compositional patterning observed here is expected to influence dislocation-mediated plasticity as well as grain boundary-mediated mechanisms such as migration and sliding. Dislocation behavior is highly sensitive to LCO [15,93,94], and the ordered zones identified in this work create structurally and chemically distinct regions in the immediate vicinity of the boundary. For example, Cao et al. [95] showed that the yield strength of CrCoNi varies inversely with the fraction of Cr-Cr pairs, since weak Cr-Cr bonding promotes nanoscale BCC-like defects that facilitate homogeneous dislocation nucleation. Enhanced Co-Cr ordering in LCO-rich regions reduces such pairs and raises the nucleation barrier, whereas chemically heterogeneous regions (e.g. a random solid solution) provide more potential nucleation sites. LCO also increases barriers to nanoscale segment slip, obstructing dislocation motion at the interface and limiting absorption. These effects indicate that the wide LCO-enriched zones formed near grain boundaries will significantly alter both the onset and subsequent propagation of dislocations in their vicinity.

In BCC alloys, the interaction between LCO and dislocations is strongly character dependent. In NbMoTaW, LCO enhances edge dislocation mobility but reduces screw mobility [31], producing strengthening primarily through the suppressed motion of screws. Breaking up LCO carries a high energetic cost, elevating the critical stress for screw motion and promoting pinning mechanisms such as cross-slip locking, while the locally flattened energy landscape associated with local chemical uniformity can reduce solute drag and facilitate edge motion. As a result, significant anisotropy in dislocation behavior is expected within LCO-rich regions adjacent to BCC grain boundaries.

These regions can also have direct implications for grain boundary mobility. As grain boundaries migrate, they must overcome both the energetic barriers associated with structure shuffling at the interface and the additional resistance created by the ordered compositional waves extending into the grain. This chemical drag effect would be particularly pronounced at lower temperatures, where the ordering is more stable and penetrates deeper into the grain, requiring disruption of this extended ordered zone during migration and leading to sluggish grain growth kinetics. Such constraints are expected to slow boundary motion, particularly under conditions where diffusion can keep pace with interface migration and atoms can relax into new low-energy configurations after the boundary passes. Furthermore, this mechanism differs from traditional solute drag, where individual segregated atoms create localized energetic barriers. Instead, the collective rearrangement of ordered domains during boundary motion could create a more substantial kinetic barrier, especially considering the strong ordering observed adjacent to the interface. Such ordering-induced drag effects could contribute to the exceptional thermal stability reported in some CCAs [22–24].

It is important to emphasize that this study does not focus on a single interatomic potential, boundary geometry, or alloy class, and therefore the underlying physical principles are broadly applicable and relevant to other CCAs and grain boundary configurations. Building on the present work and recent studies [13,69], we hypothesize that chemical transitions near interfaces are a common, if not ubiquitous, feature in CCAs exhibiting strong chemical ordering tendencies coupled with some degree of grain boundary segregation. In other words, if one is investigating the potential LCO for a given alloy, they should focus their characterization efforts on grain boundaries, especially those whose boundary planes align appropriately with candidate ordering motifs. Thermodynamic stability of the boundary structure is also an essential consideration, since any structural transition that modifies the atomic configuration is expected to alter segregation behavior and, in turn, the development of near-boundary LCO. Moreover, strategic selection of grain boundary geometry could enable tailored compositional patterns. For example, coherent twin boundaries with minimal segregation and structural disorder should remain chemically homogeneous, as demonstrated by Li et al. [18]. In contrast, high-angle boundaries with strong segregation and aligned along preferential ordering planes may significantly amplify compositional patterns. Exploring mixed boundaries provides an efficient approach to sampling a variety of ordering vectors, potentially producing long-range, anisotropic compositional patterns such as those observed in cermets [96] and a Pt-3 at.% Ni alloy [87].

## IV. SUMMARY AND CONCLUSIONS

This study used atomistic simulations of FCC and BCC CCAs to reveal how grain boundary segregation and defect structure influence the amplification of near-boundary compositional patterns and LCO. By examining multiple crystal structures, boundary types, and interatomic potentials, the findings are broadly applicable to CCAs with strong ordering tendencies. The key conclusions are summarized as follows:

1) In both CrCoNi and NbMoTaW, a dominant segregant enriches the grain boundary. This segregation locally reduces chemical complexity, creating a chemically distinct interface that serves as a template for the amplification and orientation of LCO in nearby regions.
2) In CrCoNi, periodic Ni enrichment and depletion emerge near the Σ11 grain boundary, accompanied by antiphase oscillations of Co and Cr along a direction normal to the grain boundary plane. Ni-depletion promotes symmetric Co-Cr ordering near the interface.

These oscillations maintain the same wavelength/size as isolated LCO within the grain interior, suggesting a common ordering mechanism. The magnitude and spatial extent of these patterns are temperature-dependent, with higher temperatures yielding only a single Ni-depleted peak and relaxation chemical patterning.

3) In NbMoTaW, alternating Mo- and Ta-enriched planes develop near the Σ3 (221)/(001) boundary. This ordering is driven by Nb segregation and enhanced Mo-Nb interactions resulting from Ta depletion at the boundary.

4) Simulations of Σ3 models with varying inclination angles show that boundary geometry critically influences near-boundary ordering as it provides a template for LCO. Rotating away from the optimal boundary plane weakens the LCO signal, making compositional patterning weaker. Tailoring composition, particularly by suppressing interfering segregation behavior, is another effective strategy to enhance LCO visibility.

5) Thicker bicrystal models confirm that LCO remains uniform along the boundary throughout the sample thickness yet becomes more difficult to identify within the bulk. The persistent ordering near the interface should enhance LCO detectability and will serve as a practical target for future characterization experiments.

In contrast to earlier work that focused primarily on identifying segregation states and their effects in CCAs [13,65,72,97,98], this study demonstrates how grain boundaries can be deliberately leveraged as structural templates to amplify and localize LCO that would otherwise remain diffuse within the grain interior. By systematically varying tunable grain boundary parameters, such as crystallographic orientation, and controlling segregation through composition, these results establish a mechanistic basis for engineering LCO at specific defects rather than treating ordering as a passive, bulk property. This perspective shifts the field from predicting segregation to actively designing boundary environments that promote ordered states. These results further suggest that experimental efforts to characterize LCO should focus on the compositional variations induced by segregation and structural relaxation at and near defects. The strong amplification of LCO at grain boundaries provides a potential pathway for detecting LCO in future experiments, owing to the well-defined ordering that develops uniformly through the sample thickness. More broadly, this study provides mechanistic insight into how segregation and boundary structure can localize and amplify LCO, with implications for tailoring defect chemistry, stabilizing interfaces, and ultimately influencing properties in these alloys.

## ACKNOWLEDGMENTS

This paper was supported by the National Science Foundation Materials Research Science and Engineering Center program through the UC Irvine Center for Complex and Active Materials (DMR-2011967).

## Supplementary Material

## Supplementary Note 1

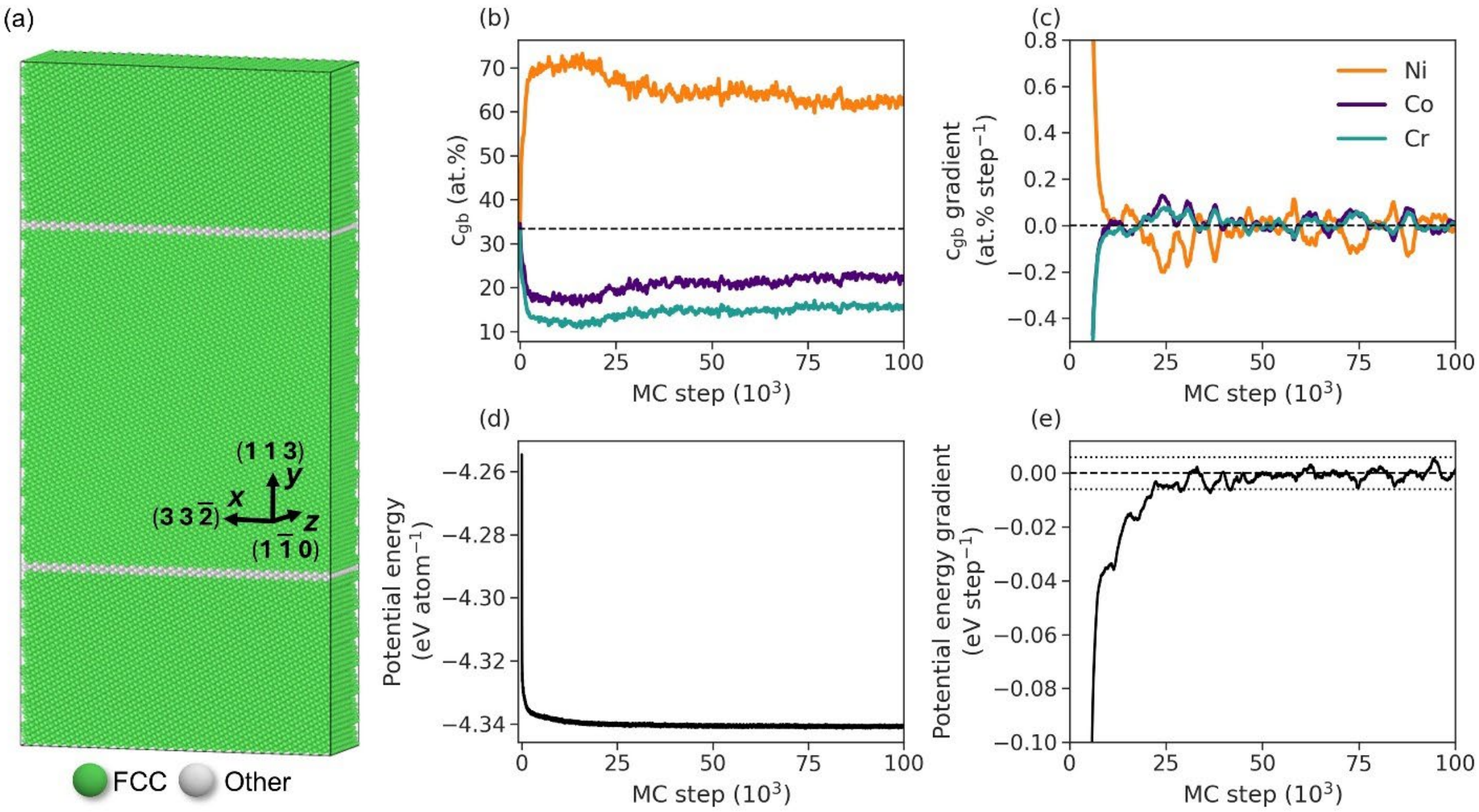

**FIGURE S1. Atomistic model and convergence study. (a) A snapshot of the biocrystal with atoms colored by structure type, where green represents FCC atoms and white represents grain boundary atoms. (b) Grain boundary concentrations of each element plotted as a function of MC step at 500 K, and (c) the gradients of each compositional slope taken over a 5,000 MC step window. (d) Potential energy per atom plotted by MC step, and (e) the respective gradient of potential energy with dotted lines denoting +/- 6 meV/step.**

Interface concentrations and potential energy for CrCoNi samples were monitored to evaluate chemical and structural convergence, as shown in Figure S1. While grain boundary composition (Figure S1(b)) and potential energy per atom (Figure S1(d)) appeared stable after approximately 30,000 MC steps, their gradients (Figures S1(c) and (e)), calculated as a moving average over 5,000 steps, revealed significant fluctuations over longer simulation periods. The simulations were determined to have converged when the absolute value of the total potential energy gradient dropped below 6 meV step$^{-1}$ (dotted lines in Figure S1(e)). At this point, both the energy and grain boundary concentrations plateaued, exhibiting only minor fluctuations.

**Supplementary Note 2**

To ensure that compositional oscillations for CrCoNi described in the main text are not an artifact of the chosen potential, a second set of limited simulations was performed using a new yet inefficient potential. A known limitation of the EAM potential used for the majority of this work is its inability to exactly replicate all pairwise chemical trends from DFT calculations. For example, Tamm et al. [1] and Ding et al. [2] used DFT-MC simulations to demonstrate significant Co-Cr ordering at 500 K and identified an attractive interaction between Ni and Cr. These attractions are attributed to the energetic penalty of like-spin Cr-Cr nearest neighbors, which likely plays a critical role in defining the alloy's dominant bonding preferences. Specifically, antiferromagnetic frustration among Cr atoms reduces Cr clustering, encouraging greater Ni and Co occupancy in Cr's coordination shell [3]. However, the EAM potential developed by Li et al. [4] predicts a repulsive interaction between Ni and Cr. Consequently, a natural question is whether near-boundary behavior changes when such chemical interactions are accounted for. To address this, atomistic simulations of single-crystal and bicrystalline models were performed using a recently developed neural network potential (NNP) [5,6] designed to more accurately model LCO in CrCoNi. The NNP's flexible architecture allows it to represent the complex potential energy surface of a multicomponent alloy when trained on a high-quality DFT dataset. Unlike the EAM, the NNP captures attractive Cr-Ni and Cr-Co interactions, as well as Cr-Cr repulsion, in agreement with DFT-based MC simulations and experimental data [1,2,7,8]. Additionally, it reproduces superlattice chemical ordering patterns analogous to those experimentally observed by Zhou et al [7]. These patterns feature alternating Cr-enriched {113} and {110} planes interspersed with Co and/or Ni-enriched planes. However, it is essential to note that the NNP also has practical limitations due to the increased complexity of its functional form for atomic interactions compared to the EAM potential. This added complexity reduces computational efficiency, necessitating significantly smaller simulation cell sizes when modeling the defect environment. Simply put, the wide variety of temperatures and segregation states reported on in the earlier parts of this manuscript would be intractable with this NNP in a reasonable amount of time.

Given the strong influence of chemical ordering on the near-boundary segregation in the EAM potential, an initial comparison of these parameters between the two potentials was performed. First, atom-swap based MC/MD simulations were conducted on a 4000-atom single-crystal cube relaxed at 200 K using the NNP with an MD timestep of 2.5 fs. Every 2 MD steps, one MC step was performed, with atom swaps attempted for 4% of the atoms. Figure S2(a) shows the Warren-Cowley parameter that describes bulk LCO tendencies for each nearest-neighbor chemical pair interaction for both the EAM potential and the NNP, where notable differences can be identified for the Ni-Ni, Ni-Cr, Co-Co, and Co-Cr interactions. For example, the strongly attractive Ni-Ni and Co-Cr interactions predicted by the EAM potential are reduced by more than half in the NNP, while the repulsive Ni-Cr and Co-Co interactions in the EAM are reversed to become attractive in the NNP. Importantly, the Ni-Cr interaction predicted by the NNP demonstrates better quantitative agreement with prior DFT studies [1,2].

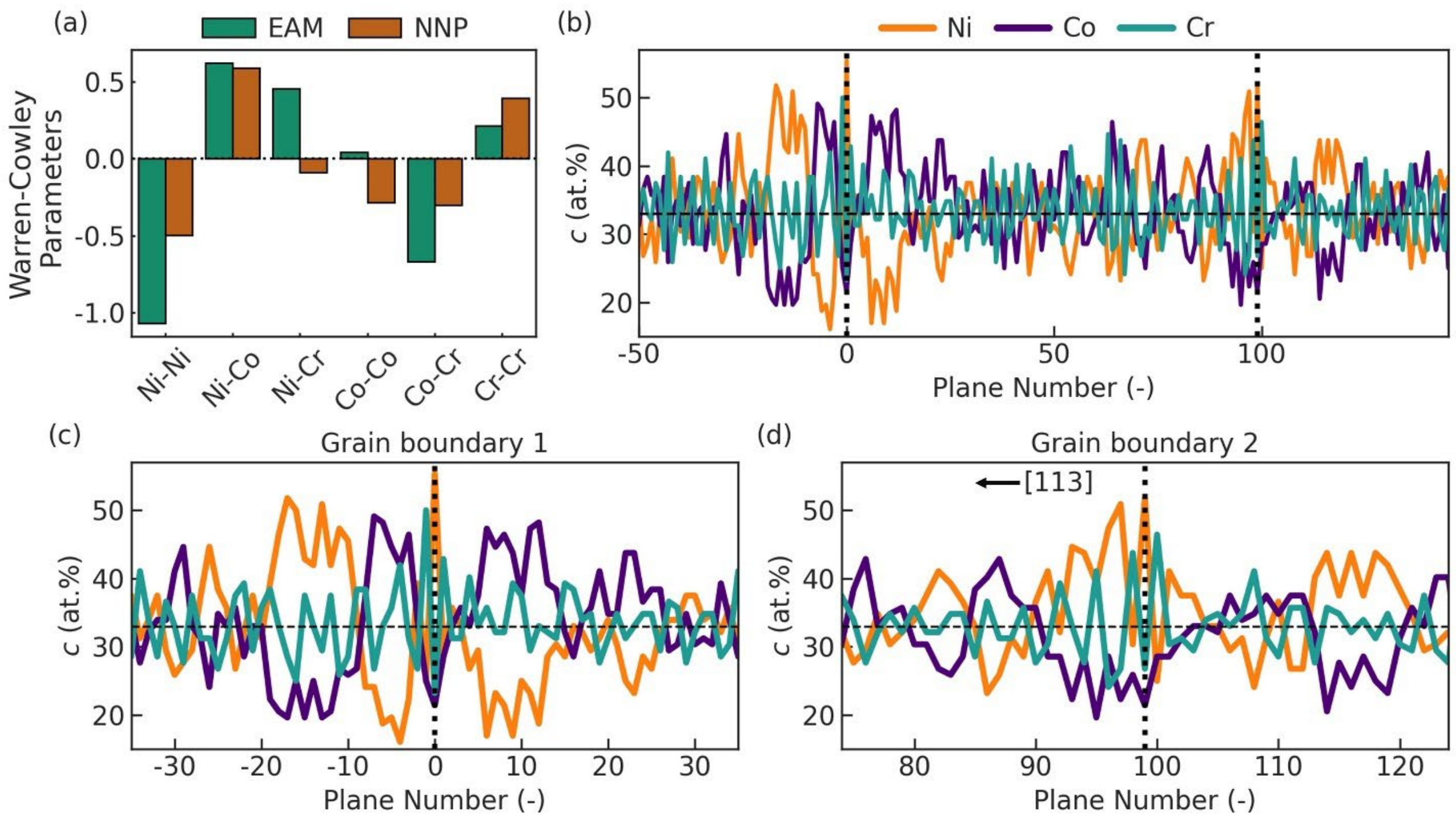

**FIGURE S2. Chemical analysis of single-crystalline and bicrystalline models relaxed using the NNP after 1000 steps. (a) Comparisons of the Warren-Cowley parameter between the EAM potential and the NNP. (b) Elemental concentrations for a sample at 200 K, with dotted vertical lines highlighting the central grain boundary planes. (c-d) Zoomed-in view spanning 6-7 nm at and near the left-hand and right-hand grain boundaries, respectively.**

Next, MC/MD simulations were performed on bicrystalline models containing two symmetric Σ11 grain boundaries at 200 K. Due to the high computational cost of the NNP, this low-temperature condition was chosen as the most practical setting to efficiently observe LCO behavior. The low speed of the potential also explains why a broad temperature study, such as that performed using the EAM potential, is not feasible with the NNP, as will be further discussed in the following section. To accommodate the extended simulation times required for full relaxation, the simulation cell dimensions were reduced to approximately $L_x \sim 3$ nm, $L_y \sim 21$ nm, and $L_z \sim 3$ nm, comprising only 22,176 atoms. The elongated *y*-dimension minimized grain boundary interactions and allowed chemical ordering to develop parallel to the interfaces. Figure S2(b) presents compositional profiles across {113} planes parallel to the grain boundary after 1000 MC/MD steps. Dotted vertical lines in Figure S2(b) denote the positions of the two grain boundaries, with Ni enrichment localized at the central grain boundary plane and Cr enrichment at adjacent planes. The small cross-sectional area of the simulation cell introduced compositional noise, manifesting as moderate elemental fluctuations throughout the sample between 25 and 41 at.%. Despite this, near-boundary compositional patterns are evident in Figures S2(c) and (d). In Figure S2(c), Ni depletion (~16 at.%) and Co enrichment (~49 at.%) occur on both sides of the boundary, similar to trends from simulations with the EAM potential. Notably, a wave-like pattern of alternating Ni and Co emerges on the left side of the boundary. At the second boundary (Figure S2(d)), the left side of the interface exhibits alternating Ni and Cr patterns consistent with the {113} plane ordering identified by Zhou et al. [7], with Cr peaks flanked by two layers of Ni enrichment. This pattern is amplified by the defect region but is asymmetric about the grain boundary plane. On the right side, Ni and Co waves are visible, though their clarity is reduced in part due to the compositional noise from the sample size. The concentration wave prominence may also depend on the symmetry across the boundary; when the symmetry is broken, the waves are less distinct and may have reduced magnitudes.

During the hybrid MC/MD simulations using the NNP, the average potential energy per atom was monitored (Figure S3). After 1000 steps, the potential energy decreases significantly, reflecting the relaxation achieved through the MC/MD procedure. Anlaysis of the compositional profiles at this stage (Figure S2) reveals the formation of near-boundary compositional patterning without significant bulk oscillations. The emergence of these near-boundary patterns after just 1000 steps highlights a strong driving force for interface-driven ordering in adjacent crystalline sites, even as LCO within the bulk continues to develop.

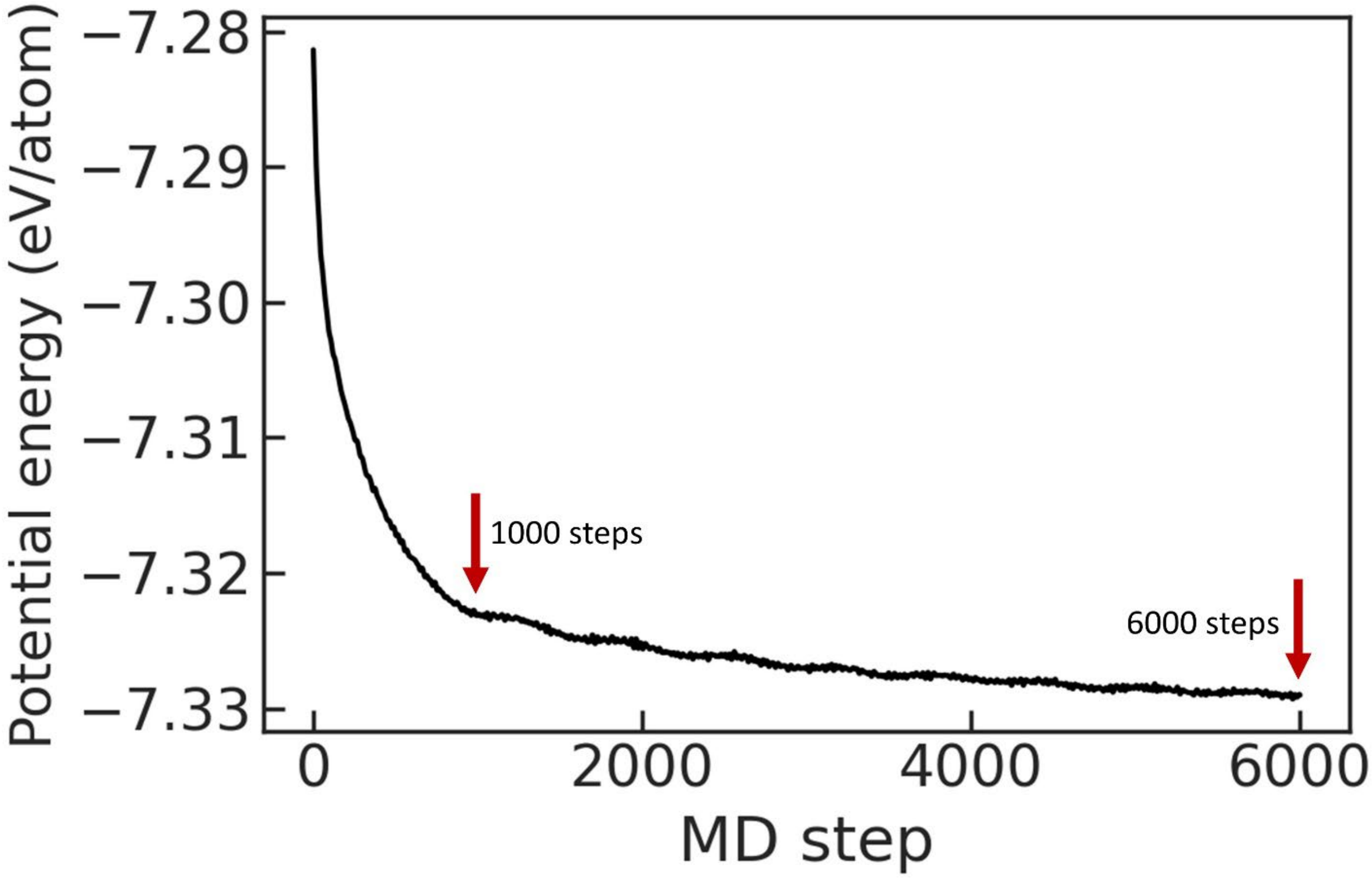

**Fig. S3. Average potential energy per atom as a function of MD step for a bicrystalline sample at 200 K. Red arrows indicate approximate regions on the curve where atomic snapshots are analyzed both in the main text and in the supplementary materials.**

Additional MC/MD steps (6000 total) show further reduction of potential energy (Figure S3). A snapshot of the planar compositions after 6000 steps is shown in Figure S4. While the near-boundary compositional patterning remains largely consistent with those observed at 1000 steps, the bulk peaks become more pronounced, diminishing the relative prominence of structure-induced fluctuations near the interface. These enhanced bulk peaks are primarily attributed to the small simulation size, which amplifies nanoscale chemical ordering even in the absence of a grain boundary acting as a structural template. Increasing the cross-sectional area beyond the characteristic ordering scale would likely suppress these bulk peaks, as overlapping ordered clusters would average out, reducing their prominence as shown in Figure 4 of the main text. However, conducting simulations with a sufficiently large cross-sectional area is computationally prohibitive due to the inefficiency of the interatomic potential.

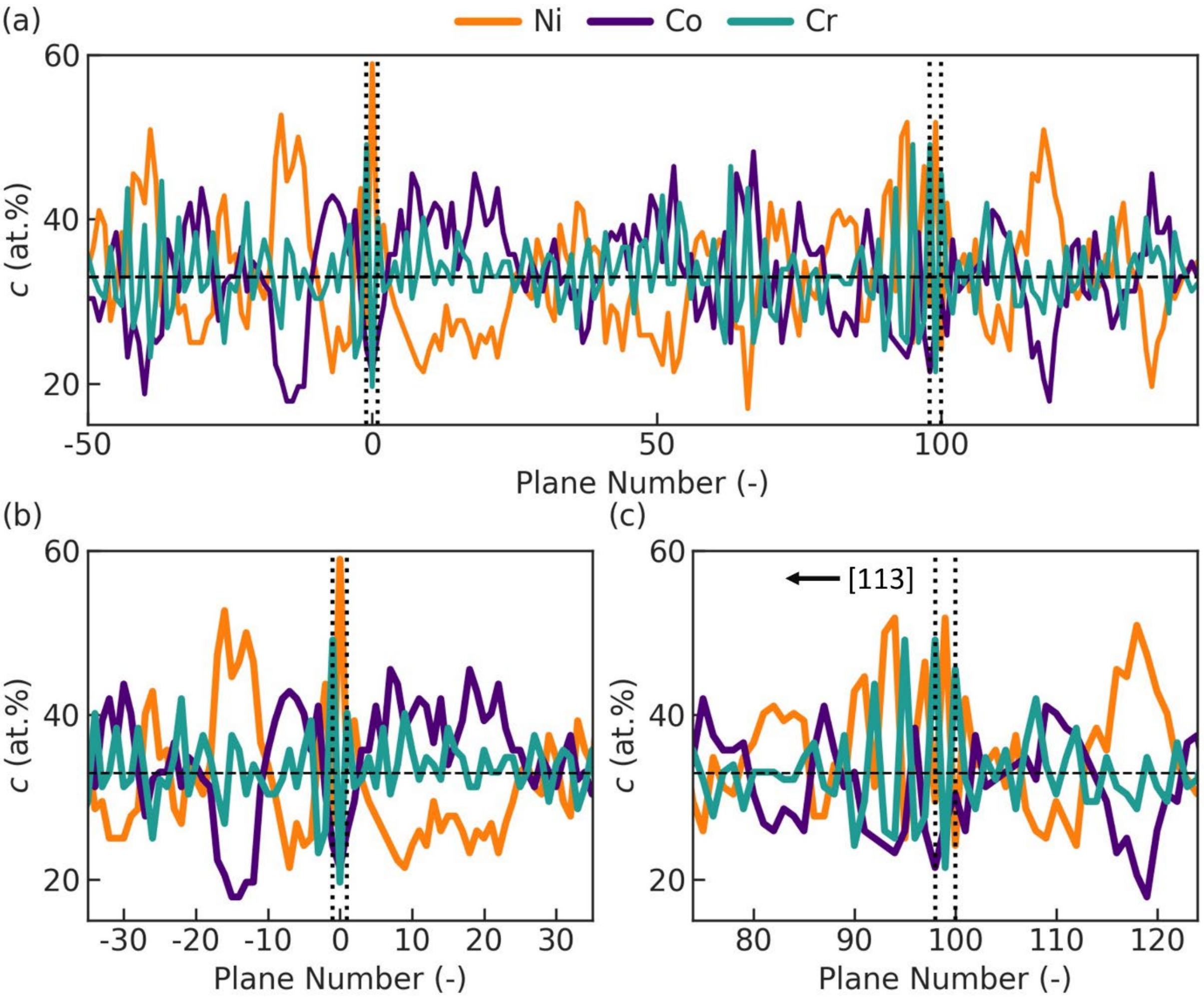

**Fig. S4. Local elemental compositions averaged over planes parallel to the grain boundary after 6000 steps. (a) Elemental concentrations for a sample at 200 K, with dotted vertical lines highlighting the central grain boundary planes. (b-c) Zoomed in view at and near the left-hand and right-hand grain boundaries, respectively.**

It is important to emphasize that a key source of the differences between the EAM and NNP results also lies in the extent of equilibration that can be practically achieved with each potential. In the NNP simulation presented here, approximately 2.6 million attempted atom swaps were performed over 6000 MC/MD steps, requiring roughly nine days of computation with significant dedicated resources. In contrast, comparable EAM simulations reached or exceed 0.5-1 billion attempted swaps, depending on system size and temperature, with each simulation typically completing within a day. Achieving similar sampling with the NNP would require 5-10 years of continuous computation for each calculation, highlighting the method's computational intractability for long-timescale equilibration. We also note that the need for multiple simulation temperatures, simulation sizes, and simulation runs to ensure repeatability would make this even more of a burden. Despite these limitations, the NNP still captures key features predicted by the EAM, including the emergence of grain boundary segregation patterns. The qualitative agreement suggests that these patterns reflect a favorable energetic state that would continue to develop with extended sampling, rather than being artifacts of any one interatomic potential.

**Supplementary Note 3**

Additional simulations of a high-angle boundary were performed for the CrCoNi alloy to help clarify the structural features that promote or inhibit extended ordering. The boundary selected was a [1-10] symmetric grain boundary rotated by 136°, which terminates on {775} planes. The atomic structure of the boundary is shown in Fig. S5(a), where the interface atoms are noticeably more disordered than the Σ11 boundary, lacking a repeatable structural motif and containing many HCP-coordinated atoms. The corresponding compositional maps (Fig. S5(b)), plotted by plane with the grain boundary center marked by dashed lines, show clear Ni enrichment and Co/Cr depletion at the interface, consistent with the Σ11 boundary. Ni-depleted regions immediately adjacent to the grain boundaries are also observed, which come from moderate Ni interfacial segregation that facilitates anti-phase Co-Cr fluctuations. Although the fluctuations qualitatively resemble the Σ11 case, the magnitudes of the peaks are substantially reduced. For example, near-boundary Co-Cr peaks adjacent to the Σ11 approach 50 at.% (only ~40 at.% here) while Ni concentration reaches 0-10 at.% (~20 at.% here). As a result of these milder fluctuations, the spatial extent of the waves into the grain is also reduced. The residual fluctuations that do extend into the crystal are comparable in amplitude to random variations within the grain interior, indicating that experimental characterization of this boundary would reveal only weak LCO signals that barely exceed background noise. These results suggest that Ni segregation alone can provide some localization of ordering, but that strong amplification requires a uniform and symmetric interfacial motif like that of the Σ11 boundary. This highly ordered grain boundary structure allows for interactions to be more uniform along the planes adjacent to the boundary, as is the case for the BCC alloy. While other boundaries will likely meet this criterion, a comprehensive survey of many FCC boundary types falls outside the scope of the present work, and we instead chose to focus on broader applicability across alloy classes.

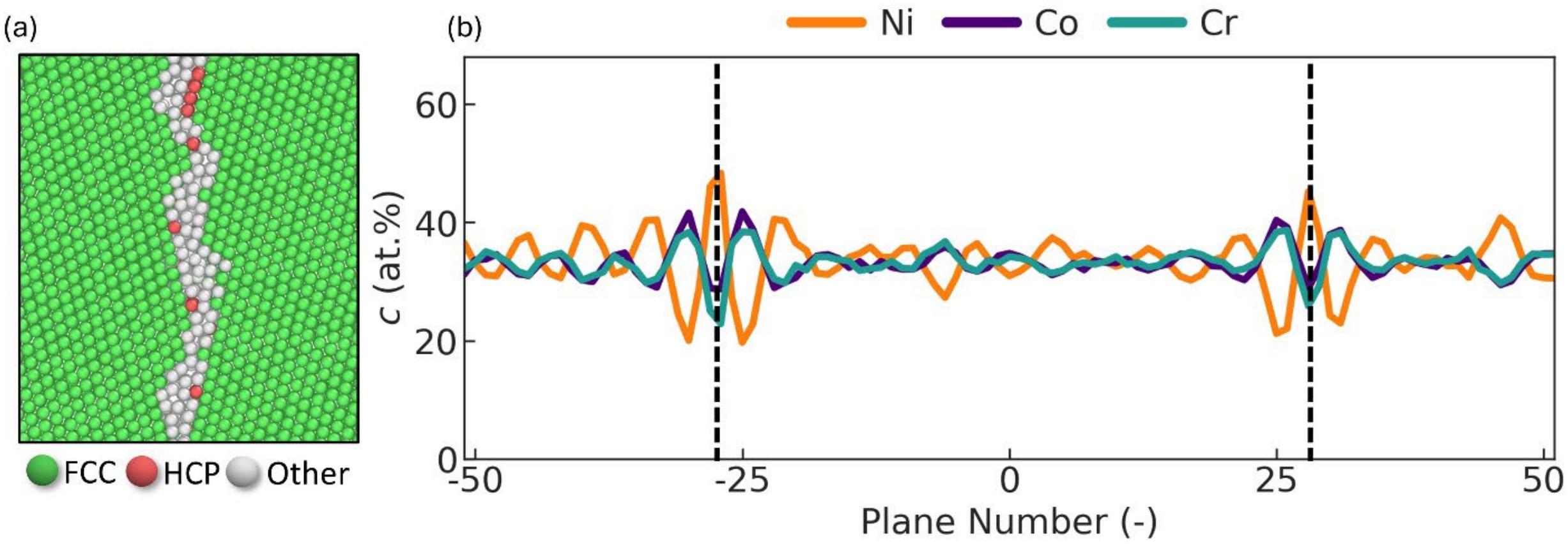

**FIGURE S5. Structure and compositional profile for a symmetric [1-10] boundary rotated by 136°. (a) The**

**grain boundary structure. (b) The compositional profile across the entire simulation cell, where the dashed lines indicate the center of the grain boundary plane.**

## Supplementary Note 4

To examine whether the behavior observed at the asymmetric Σ3 boundary is unique, we performed additional simulations of Nb10Mo25Ta25W40 at a symmetric Σ27(552)[1-10] boundary (Fig. S6(a)), which is a high-angle boundary and a misorientation previously investigated in pure BCC metals [9]. These simulations show that Nb and Mo segregate strongly to the Σ27 boundary (gray shaded region in Fig. S6(b)), consistent with the segregation observed at the Σ3. However, while Mo enrichment extends into the plane directly adjacent to boundary sites, no further amplification of LCO develops. The compositional modulations in this near-boundary region (20-27 at.%) are substantially smaller than those at one side of the Σ3 boundary, where local concentrations reach ~60 at.% and even exceed the increased defect concentrations. This indicates that although Mo enrichment near the boundary is promoted by strong Nb segregation, the absence of the appropriate terminating plane prevents the emergence of extended ordering motifs.

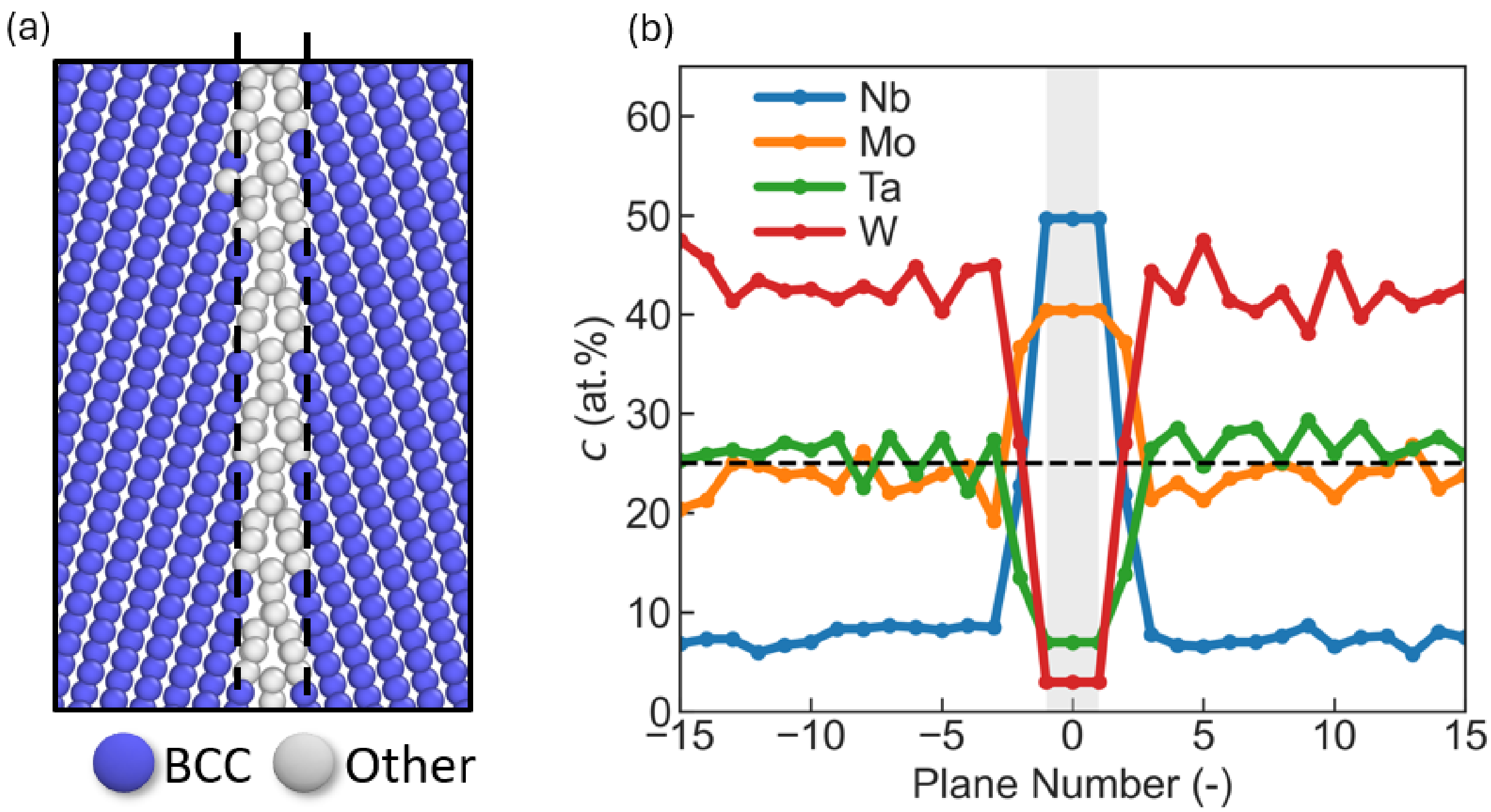

**FIGURE S6. Structure and compositional profile for a Nb10Mo25Ta25W10 alloy at a Σ27(552)[1-10] boundary. (a) The grain boundary structure. (b) The compositional profile across the boundary, where the shaded region denotes the grain boundary sites. The grain boundary concentration is averaged over the defect region.**